\documentclass[12pt]{article}
\usepackage{a4}
\usepackage{amsfonts}
\usepackage{amssymb}
\usepackage{epsfig}

\newcommand{\be}{\begin{equation}}
\newcommand{\ee}{\end{equation}}
\newcommand{\bea}{\begin{eqnarray}}
\newcommand{\eea}{\end{eqnarray}}
\newcommand{\mbb}{\mathbb}
\newcommand{\ti}{\times}
\newcommand{\half}{\frac{1}{2}}
\newcommand{\third}{\frac{1}{3}}

\newcommand{\twothirds}{\frac{2}{3}}
\newcommand{\mc}{\mathcal}

\newcommand{\K}{\mc{K}}

\begin{document}

\title{
\begin{flushright} \vspace{-2cm}
{\small DAMTP-2005-48 \\ \vspace{-0.35cm}
hep-th/0505076} \end{flushright}
\vspace{3cm}
{\bf Large-Volume Flux Compactifications: Moduli Spectrum and D3/D7
Soft Supersymmetry  Breaking}
}
\author{}
\date{}

\maketitle

\begin{center}
Joseph P. Conlon, \footnote{e-mail: J.P.Conlon@damtp.cam.ac.uk}
Fernando Quevedo \footnote{e-mail: F.Quevedo@damtp.cam.ac.uk}
and
Kerim Suruliz \footnote{e-mail: K.Suruliz@damtp.cam.ac.uk}\\
\vspace{0.5cm}
\emph{DAMTP, Centre for Mathematical Sciences,} \\
\emph{Wilberforce Road, Cambridge, CB3 0WA, UK} \\
\end{center}

\begin{abstract}
\noindent We present an explicit calculation of the spectrum of a general class of string models, corresponding
to Calabi-Yau flux compactifications with $h_{1,2}>h_{1,1}>1$ with leading perturbative and non-perturbative
corrections, in which all geometric moduli are stabilised as in hep-th/0502058. The volume is exponentially
large, leading to a range of string scales from the Planck mass to the TeV scale, realising for the first time
the large extra dimensions scenario in string theory. We provide a general analysis of the relevance of
perturbative and non-perturbative effects and the regime of validity of the effective field theory. We compute
the spectrum in the moduli sector finding a hierarchy of masses depending on inverse powers of the volume. We
also compute soft supersymmetry breaking terms for particles living on D3 and D7 branes. We find a hierarchy of
soft terms corresponding to `volume dominated' F-term supersymmetry breaking. F-terms for K\"ahler moduli
dominate both those for dilaton and complex structure moduli and $D$-terms or other de Sitter lifting terms.
This is the first class of string models in which soft supersymmetry breaking terms are computed after fixing
all geometric moduli. We outline several possible applications of our results, both for cosmology and
phenomenology and point out the differences with the less generic KKLT vacua.

\end{abstract}

\thispagestyle{empty}
\clearpage

\tableofcontents

\section{Introduction}
\label{Introduction}
\linespread{1.2}

We are entering a new phase in string phenomenology. There now exist well controlled mechanisms to stabilise the
moduli fields and break supersymmetry, the two outstanding problems
for low-energy string models
\cite{fluxes, hepth0105097,hepth0301240}. Furthermore
quasi-realistic models have been constructed that fit precisely within this moduli-fixing scenario. For the
first time we can do proper phenomenology, starting with the computation of the spectrum of low-energy particles
after having found a well-defined vacuum solution with all moduli stabilised. This programme has not been fully
attempted so far. The main obstacle has been the lack of a concrete case in which the spectrum and soft
supersymmetry breaking terms can actually be computed after
stabilising all moduli.

Having a concrete model with all moduli fixed allows, first of all, the
computation of all relevant scales from first principles: the string
scale, Kaluza-Klein masses, gravitino mass, as well as masses of the
different particles in the moduli sector. Furthermore, it allows the
computation of the magnitude of soft supersymmetry breaking terms in the observable
sector of the theory.

There have been several important attempts at computing soft supersymmetry breaking terms from flux
compactifications in type IIB string theory \cite{hepth0311241, hepth0408036, hepth031232, hepth0406092}. 
However, not all geometric moduli are fixed by the fluxes. 
The standard procedure to fix the remaining moduli, namely the
KKLT scenario which involves the introduction of non-perturbative effects from D7 branes or D3 instantons, actually
restores supersymmetry in an AdS minimum. Then in this scenario the breaking of supersymmetry is entirely due to
the mechanism that lifts the minimum to de Sitter space, thus erasing the effects of fluxes in the calculation
of the spectrum of moduli and matter fields. Recently there have been attempts to address this problem
\cite{hepth0411066, ciqs, choi, hepph0504036}. 
The main obstacle here is  the lack of simple concrete realisations of the KKLT scenario in
which soft breaking terms can be computed. In particular, in explicit models, what was a minimum in the complex
structure and dilaton directions may become a saddle point in the
potential after including the K\"ahler moduli
\cite{hepth0411066, ciqs, hepth0404116} .

Nevertheless, this is only a manifestation of the model dependence of the KKLT scenario. There should be many
models with varying numbers of complex structure moduli in which a
concrete KKLT minimum can be found - for the state of the art see
\cite{hepth0503124}. However,
explicit control of the effective potential dependence on all the moduli tends to be lost once there are more
than a few complex structure moduli. The lack of concrete calculations in this case is only a human limitation
rather than a problem with the KKLT minimum.

Fortunately a large new class of minima of the full scalar potential, differing from the KKLT solutions,  has
been recently uncovered \cite{hepth0502058} (also see
\cite{hepth0408054}). 
The main  ingredient here is the realisation that after including
both perturbative and non-perturbative corrections to the scalar
potential, a very general minimum emerges 
which is non-tachyonic in 
all directions in geometric moduli space. General arguments show that
this occurs for all Calabi-Yau
compactifications with $h^{1,2}>h^{1,1}>1$. A simple example with two K\"ahler moduli, the
${\mathbb{P}}^4_{[1,1,1,6,9]}$ model, was studied in detail and the minimum found explicitly. An important
property of this minimum is that it is at exponentially large
volume. This creates a naturally small expansion
parameter, namely the inverse volume, which allows good control of otherwise
intractable calculations. Having such a large volume 
gives rise to a stringy realisation of the large extra dimensions
scenario \cite{add}.
It is then natural to study in detail this class of models and in particular the
explicit two K\"ahler moduli case.

There are several important properties
 of these new vacua. They are generic and the corresponding field theory is valid
for a much broader range of parameters than the KKLT
 minima. Furthermore, 
they lack the many tachyonic directions
that often appear in the KKLT solutions and are thus more robust than the corresponding KKLT
vacua. They are also physically very different. Besides the much larger volume, the $AdS$ minimum is
non-supersymmetric and therefore the sources of supersymmetry breaking are many: fluxes, non-perturbative
effects and $dS$ lifting mechanisms. In KKLT the $AdS$ minimum is supersymmetric and the sole source of
supersymmetry breaking is the $dS$ lifting mechanism.

The purpose of this article is to start a complete phenomenological analysis of these models. We first compute
the spectrum in the moduli sector as a function of the model parameters, in particular the volume. We obtain the
volume dependence of the various mass scales present, and find a natural hierarchy between the string scale,
Kaluza-Klein masses, gravitino mass, complex structure-dilaton masses
and K\"ahler moduli masses. The mass
scales depend on different inverse powers of the volume except for the axionic partner of the `volume' modulus,
whose mass is an inverse exponential of the volume and thus extremely small.  We find that exponentially large
volumes are generic and for $\mc{O}(1)$ values of the parameters (fluxes, terms in the non-perturbative
superpotential, etc.) a wide range of string scales can be explicitly obtained, from the Planck mass to the
electroweak and beyond. We also perform the corresponding analysis for the soft breaking terms for matter on
D3/D7 branes and find a hierarchy of soft terms dominated by the $F$-term of the volume modulus.

The article is organised as follows. In section 2 we give a brief overview of flux compactifications of type IIB
string theory. In section 3 we present a general discussion of the limitations of the standard KKLT effective
potential, caused by the neglect of the perturbative corrections to
the K\"ahler potential. 
This restricts the volume to be not too large and the range
of values of the flux superpotential to be extremely small. We point
out that if the leading $\alpha'^3$
corrections are included, the potential can be trusted for a far
broader range of values. 
The robustness of the known $\alpha'^3$ correction
is critically investigated and we conclude that possible extra bulk corrections are further suppressed by
negative powers of the volume. Section 4 reviews the class of models we consider and provides a detailed
calculation of the magnitude of the relevant scales (string scale, Kaluza-Klein scale, gravitino mass, moduli
and modulino masses). We explicitly compute the spectrum for the two K\"ahler moduli Calabi-Yau
$\mbb{P}^4_{[1,1,1,6,9]}$ mentioned above. We also present results for the general case, illustrating our
conclusions with a three K\"ahler moduli example (the ${\cal{F}}_{11}$ model of \cite{hepth0404257}).

Section 5 concentrates on the soft supersymmetry breaking terms for both D3 and D7 matter fields. We find a
hierarchy of supersymmetry breaking in which the $F$-term corresponding to the volume modulus dominates. The
effects of $D$-terms or IASD fluxes for the de Sitter lift are also discussed.
We conclude the article with a general comparison with the KKLT
scenario, pointing out the similarities and differences. Finally we
present our conclusions and general outlook.

\section{Flux Compactifications of IIB String Theory}
\linespread{1.2}

We shall first briefly review flux compactifications of type IIB
string theory mostly to establish conventions and to collect the basic
equations we will be using and modifying later.

The ten-dimensional bosonic  massless fields consist of the metric $g_{MN}$, scalars $\phi,
C_{0}$, RR antisymmetric tensors $C_{2}$, $C_{4}$, the latter with
self-dual field strength, and the NS antisymmetric tensor $B_2$.
To obtain four-dimensional $\mc {N}=1$ models, we compactify on
Calabi-Yau orientifolds. Fluxes for the RR three-form
$F_3= d C_2$
and NS three-form $H_3= dB_2$ can be turned on, and must satisfy the
standard quantisation conditions:
\be
\label{quantisation}
\frac{1}{(2 \pi)^2 \alpha'} \int_{\Sigma_a} F_3 = n_a \in \mbb{Z}, \quad \quad \frac{1}{(2 \pi)^2 \alpha'}
\int_{\Sigma_b} H_3 = m_b \in \mbb{Z},
\ee
where $\Sigma_{a,b}$  represent three-cycles of the Calabi-Yau
manifold. The metric is then a warped product of flat 4-dimensional
spacetime and a conformally Calabi-Yau orientifold, with the five-form field
strength also dependent on the warp factor.

This compactification generically has $O3/O7$ orientifold
planes, D3/D7 branes and fluxes. These all contribute to the $C_4$
tadpole that must be cancelled. This condition reads:
\be
\label{tadpolecancellation}
N_{D3} - N_{\bar{D}3} +  \frac{1}{(2\pi)^4 \alpha'^2} \int H_3 \wedge F_3 = \frac{\chi(X)}{24}.
\ee
Here $\chi(X)$ collects the contribution to D3 brane charge 
from orientifold planes and D7 branes. In the F-theory
interpretation,
$\chi(X)$ is the Euler number of the corresponding
four-fold.
Gauge theories live on  the world-volume of both D3 and D7
branes, and can give rise to either standard model or hidden sector matter.

The effective field theory corresponds to a standard  $\mc{N}=1$
supergravity theory, with the superpotential being of Gukov-Vafa-Witten
type \cite{hepth9906070}:
\be
\label{GVW}
W = \int_M G_3 \wedge \Omega,
\ee
where $G_3 = F_3 - i S H_3$, with $S$ \footnote{We shall denote the
  dilaton-axion and K\"ahler moduli by $S$ and $T$ respectively. These
  are related to the $\tau$ and $\rho$ frequently used by $\tau = iS$
  and $\rho = i T$.} being the dilaton-axion
\be
iS \equiv \tau =  C_0 + i e^{-\phi},
\ee
 and
$\Omega$ the holomorphic $\left(3,0\right)$ form of the Calabi-Yau.
This superpotential depends both on the dilaton, as it appears explicitly
in the definition of $G_3$, and the complex structure
moduli $U$, as these moduli measure the size of the 3-cycles that
appear in the quantisation condition (\ref{quantisation}). 
However, $W$ does not depend
on the K\"ahler moduli.

The K\"ahler potential $\mc K$ is a sum of terms depending on the
different moduli $\mc{K}= \mc{K}_{T_i} + \mc{K}_{U_a} 
+ \mc{K}_{S}$ \footnote{We will usually denote the complex structure part of the
  Ka\"hler potential by $\mc{K}_{cs}\equiv \mc{K}_{U_a}$.}
and takes the standard form (neglecting
warp factors):
\be
\label{noscalekp}
\mc{K} =
-2 \log \left[\mc{V} \right] -
\log\left[-i \int_M \Omega \wedge \bar{\Omega}\right] - \log\left(S + \bar{S}\right),
\ee
where $\cal V$ is the classical volume of the Calabi-Yau manifold $M$ in units of $l_s = 2
\pi \sqrt{\alpha'}$:
\be
\mc{V} = \int_M J^3 = {1\over 6} \kappa_{ijk} t^i t^j t^k.
\ee
Here $J$ represents the K\"ahler class, and $t_i$, $i=1, \ldots ,h_{1,1}$
are moduli measuring the size of two-cycles. The corresponding
four-cycle moduli $\tau_i$ are defined by:

\be
\tau_i = \partial_{t_i} {\cal V} = {1\over 2} \kappa_{ijk} t^j t^k.
\ee
The complexified K\"ahler moduli are: \be T_i \equiv -i\rho_i \equiv \tau_i + i b_i, \ee with the axionic fields
$b_i$ coming from the RR four-form. The K\"ahler potential $\K_{T_i}$
is of no-scale type, with  
$G^{i\bar{\jmath}} \K_i
\K_{\bar{\jmath}} = 3$. Here we denote the K\"ahler metric by
$G_{i\bar{\jmath}} \equiv \mc{K}_{i\bar{\jmath}}=
\partial^2 \mc{K}/\partial_i \phi \partial_{\bar{\jmath}} \phi$ and  $G^{i\bar{\jmath}}=G^{-1}_{i\bar{\jmath}}$.
 Using this and the
superpotential's indepedence of the K\"ahler moduli, 
it follows that the standard $\mc{N}=1$ supergravity scalar potential:
\be
\label{Vsugra}
V = e^{\K} \left[G^{i \bar {\jmath}} D_i W \bar{D}_{\jmath} \bar{W} - 3 \vert W \vert^2 \right],
\ee
with $i,j$ running over all moduli,
becomes
\be
V_{no-scale} = e^{\K} G^{a\bar b}D_a W \bar{D}_b \bar{W},
\ee
where $a$ and $b$ run over dilaton and complex-structure moduli only.
As $V_{no-scale}$ is positive definite, we can locate the
complex structure moduli at a minimum of the potential by solving
\be
\label{csstabilisation}
D_a W \equiv \partial_a W + (\partial_a K) W = 0.
\ee
This can be done for generic choices of the fluxes and
we denote the value of $W$ following this step as $W_0$.

This procedure fixes the complex structure moduli and the dilaton but
leave the K\"ahler moduli undetermined. The dilaton and complex
structure moduli are then integrated out to focus on an effective
theory for the K\"ahler moduli. To also stabilise
these fields non-perturbative effects have to be included. These
effects
mostly arise from the gauge theories living on the D7 branes. 
For gauge fields on a D7 brane wrapping a four-cycle of size $\tau_i$,
the gauge kinetic function is
\be
f_i = \frac{T_i}{2\pi}.
\ee
Similarly the gauge kinetic function for D3 brane gauge
fields takes the universal form
\be
f= \frac{S}{2\pi}. 
\ee
Either Euclidean D3 brane instantons \cite{wittenw} or D7 brane
gaugino condensation \cite{gc}
from an anomaly free gauge theory will naturally give rise to a
non-perturbative superpotential.

Thus the full non-perturbative superpotential is expected to take
the form
\be
\label{Tsuperpotential}
W= W_0 + \sum_i A_i\,  e^{-a_i T_i},
\ee
where $A_i, a_i$ are model dependent constants.
Substituting (\ref{Tsuperpotential}) into the scalar potential
(\ref{Vsugra}) generates a
nontrivial minimum for the K\"ahler moduli, corresponding to the
solution of the equations
\be
D_i W \equiv \frac{\partial W}{\partial T_i} + W \frac{\partial
  \mc{K}}{\partial T_i}= 0.
\ee
As then $D_i W =0$ for all moduli, this
corresponds to a supersymmetric solution.
The presence of the $-3|W|^2$ term in the scalar potential ensures
that this
minimum is clearly anti de Sitter.

There are several extant proposals to lift the minimum to a de Sitter vacuum.
These all essentially add a positive definite source to the
scalar potentials. The original proposal involved anti D3 brane tensions \cite{hepth0301240},
but one could also use magnetic field fluxes on D7 branes which correspond to D-terms
\cite{hepth0309187}, or IASD fluxes generating F-terms \cite{hepth0402135}.
The values of $W_0$, as well as the lifting term, have to be carefully
chosen to give a minimum with vanishing cosmological constant.

The fixed complex structure moduli depend on the integers that
define the fluxes. There are very many discrete flux choices, 
rendering this amenable to a statistical
treatment. This has been carried out in several articles
\cite{hepth0404116, hepth0307049, hepth0411183}
(see also \cite{hepth0411173, hepth0502060}). The main results relevant
for us are that:

(i) There are an exponentially large number of solutions. 

(ii) There are regions of moduli space that act as
attractors where many solutions concentrate, namely the regions close
to conifold singularities. 

(iii) The
effective superpotential or, more properly, $e^{\K_{cs}} |W_0|^2$ is uniformly
distributed.  Thus every value is possible, up to a maximum determined by the tadpole cancellation
condition ($W_{0, max} \sim 100$), but larger values of $W_0$ are more
common than small values. 

(iv) As $T \sim \ln (W_0)$, the number of
solutions drops exponentially with the internal volume.  

(v) Finally, solutions are distributed preferentially at strong
coupling. To be more precise,
$
N(\textrm{solutions } \vert g_s < \epsilon) \sim \epsilon.
$ 

Let us note that some of these statistical results can change substantially after introducing
effects to fix the K\"ahler moduli.

\section{Perturbative  Effects in $\mc{N}=1$ Supergravity }
\label{Alpha'Section}
\linespread{1.2}

The model described in \cite{hepth0502058} and further studied  below
 involves the $\alpha'$ corrections as a crucial ingredient.
While these are always present and must be accounted for,
this invites suspicion. It is often felt that a totalitarian principle
applies: if some corrections are important, all are,
and thus it is preferable to seek vacua for which all $\alpha'$ corrections
can be neglected. We will address below the question of when this is
 possible, and show that in type IIB flux compactifications the answer
 is `almost never': the leading contribution to the scalar potential nearly always comes
from perturbative $\alpha'$ effects.
However, it is helpful to regard the type IIB case as an example of
 more general behaviour.

\subsection{General Analysis}
\label{generalanalysis}

Let us start with an $\mc{N}=1$ supergravity theory with tree-level
K\"ahler potential $\mc{K}$ and tree-level superpotential $W$. In general
$\mc{K}$ receives corrections at every order in perturbation
theory, $\mc{K}_p$, and non-perturbative corrections $\mc{K}_{np}$,
whereas $W$ is not renormalised in perturbation theory and
only receives non-perturbative corrections $W_{np}$. Therefore we can
write\footnote{The quantities $J$ and $\Omega$ introduced here should
  not be confused with the Calabi-Yau forms.}
\bea
{\K} & = & \K_0\ + \K_p + \K_{np}\approx \K_0+J, \\
W &  = &  W_0 + W_{np} \approx W_0+\Omega,
\eea
where $J$ represents the leading (perturbative) correction to $\mc{K}$ and
$\Omega$ the leading non-perturbative correction to $W$ in a coupling
expansion.
We ask when it is safe to neglect
the corrections $J$ or $\Omega$.

The F-term scalar potential is
\be
\label{FScalarPot}
V\ =\ e^{\K}\left[ D_i W\  D_{\bar k}{\bar W} \ (\K)^{-1}_{i\bar{k}}
- 3\
|W|^2 \right].
\ee
This can be expanded in powers of $J$ and $\Omega$ as follows:
\be
V\ = \ V_0 + V_J + V_\Omega + \cdots ,
\ee
where
$$
V_0 \sim W_0^2, \quad V_J \sim J W_0^2, \quad
V_\Omega \sim \Omega^2 + W_0 \Omega,
$$
and the ellipses refer to higher-order terms combining $J$ and $\Omega$.
The exact expressions for these quantities may
be explicitly computed but are not relevant for most of our argument below.

Normally, the structure of $V$ is essentially determined by $V_0$, with the other terms providing small
corrections in a weak coupling expansion. However if the tree-level potential has a flat direction along which
$V_0$ is constant, then the structure of the potential, and in particular its critical points, are determined by
the corrections. This behaviour is in fact common, with prime examples being the no-scale potentials generically
appearing in string theory. These correspond to a K\"ahler potential satisfying $G^{-1}_{i\bar{k}} \K_i
  \K_{\bar k} =3$ and a constant superpotential $W_0$. In this case
$V_0=0$ and it is the corrections that determine the
structure of the potential.
Although both $V_J$ and $V_\Omega$ will
play a role, typically we expect $V_J$ to dominate over
$V_\Omega$, as the former is perturbative in the coupling and the
latter non-perturbative.
However, since $\Omega$ is the
only correction to $W_0$ it cannot be totally neglected.

In some, very special, cases it is safe to neglect the
 $\mc{O}(J)$ corrections.
 First, if $W_0=0$ then automatically $V_J=0$ and the
leading correction comes from $V_\Omega$. For example, this occurs
in the standard heterotic string racetrack scenario.
Similarly, if $W_0 \ll 1$ in suitable units, then $\Omega$ may
be of similar magnitude to $W_0$. In this case, we have
$$
\Omega \sim W_0 \Rightarrow V_\Omega \sim \Omega^2 \textrm{ and } V_J
\sim J \Omega^2,
$$
and therefore \be \frac{V_J}{V_\Omega} \ll 1. \ee This is the relevant behaviour for the KKLT scenario. Finally,
once $W > \frac{\Omega}{J}$ (which is $ \ll 1$ as $\Omega \ll J$), the perturbative effects dominate and must be
included.

It is worth noting that the limit $W \sim \Omega$, in which the tree level superpotential is comparable to its
non-perturbative corrections, is very unnatural. There is furthermore no need to restrict to this limit if we
have information on the perturbative corrections to $\mc{K}$, which is true in both type IIB and heterotic
cases.\footnote{Note that the original proposal for
  gaugino condensaton in the heterotic string \cite{gc} included a
  constant term in the superpotential from the antisymmetric
  tensor $H_{mnp}$ as well as the nonperturbative, gaugino
  condensation superpotential for the dilaton. This was abandoned
  because the constant was found to be quantised in string units and
  could not be of order the nonperturbative correction. However,
  since the leading perturbative corrections to $K$ were found soon
  after, this avenue could have been revived in the early 90's.}

\subsection{Application to Type IIB Flux Compactifications}
\label{typeiibsection} Let us illustrate these issues in the concrete setting of type IIB flux
compactifications. As discussed above, the K\"ahler and superpotentials 
take the forms,
\bea
W & = & \frac{1}{l_s^2} \left( \int G_3 \wedge \Omega + \sum A_i e^{-
  a_i T_i} \right), \nonumber \\
\mc{K} & = & - 2\ln (\mc{V}_E) - \ln \left(-i \int \Omega \wedge
\bar{\Omega}\right) - \ln (S + S^{*}).
\eea
There is also a normalisation factor in front of $W$ that is not
important here but will be considered in section 4. 
The volume $\mc{V}_E$ and moduli $T_i$ are measured in Einstein frame
($g_{\mu \nu, E} = e^{-\frac{\phi}{2}}g_{\mu \nu, s}$). The
non-perturbative superpotential is generated by either D3-brane
instantons ($a_i = 2\pi$) or gaugino condensation ($a_i =
\frac{2 \pi}{N}$).
If the dilaton and complex structure moduli have been fixed by the fluxes, these potentials reduce to
\bea
\label{FluxPotentials}
W & = & W_0 + \sum A_i e^{- a_i T_i}, \nonumber \\
\mc{K} & = & \mc{K}_{cs} - 2\ln(\mc{V}_E).
\eea
In the language of section \ref{generalanalysis}, the scalar potential derived from
(\ref{FluxPotentials}) includes $V_\Omega$ but not $V_J$.
We now show that for almost all flux choices and moduli values, the use of (\ref{FluxPotentials})
without $\alpha'$ corrections is inconsistent. Although the argument
extends easily to any number of K\"ahler moduli, for illustration we
consider one K\"ahler modulus and the geometry appropriate to the quintic.
(\ref{FluxPotentials}) then reduces to
\bea
\label{OneModulusPotentials}
W & = & W_0 + A e^{-a T}, \nonumber \\
\mc{K} & = & \mc{K}_{cs} - 3 \ln (T+\bar{T}) \eea For the quintic $\mc{V}_E =
\frac{5}{6}t^3 = 
\frac{1}{6\sqrt{5}}(T+\bar{T})^{\frac{3}{2}} = \frac{\sqrt{2}}{3\sqrt{5}} \sigma^{\frac{3}{2}}$, where
$T = \sigma+ib$. The leading $\alpha'$ correction to the K\"ahler potential is \cite{hepth0204254} \be
\label{alpha'correction} \mc{K}_{\alpha'} = \mc{K}_{cs} - 2 \ln \left(\mc{V}_E + \frac{\xi}{2
g_s^{\frac{3}{2}}}\right), \ee where $\xi = \frac{-\chi(M) \zeta(3)}{2(2 \pi)^3} = 0.48$. The factor of
$g_s^{-\frac{3}{2}}$ arises from our working in Einstein frame; it would be absent in string frame, in which
$\mc{V}_s = \mc{V}_E g_s^{\frac{3}{2}}$. The resulting scalar potential is \be \label{StringCorrectedPotential}
V = e^{\mc{K}} \Bigg( \overbrace{\frac{4 \sigma^2 (A a)^2 e^{-2 a
      \sigma}}{3} - 4 \sigma W_0 (A a) e^{-a \sigma}}^{V_\Omega} +
      \overbrace{\frac{9 \sqrt{5} \xi W_0^2}{4 \sqrt{2} g_s^{\frac{3}{2}} \sigma^{\frac{3}{2}}}}^{V_J} \Bigg). \ee The $\alpha'$
correction $V_J$ dominates at both small and large volume. Although perhaps counter-intuitive, this is to be
expected - the $\alpha'$ correction is perturbative in volume whereas the competing terms are non-perturbative.

We may quantify when neglecting $\alpha'$ corrections is permissible. The allowed range of $\sigma$ is \be
\textrm{Max}(\mc{O}\left(\frac{1}{g_s}\right), \sigma_{min}) < \sigma < \sigma_{max}, \ee where $\sigma_{min}$ and
$\sigma_{max}$ are taken to be the solutions of $\vert V_J (\sigma) \vert = \vert V_\Omega (\sigma)
\vert$, that is \be \label{SigmaEquations} \frac{4 \sigma^2 \vert A^2 \, a^2 \vert e^{-2 a \sigma}}{3} - 4
\sigma \vert
  W_0 A \, a\vert e^{-a \sigma}
= \frac{9 \sqrt{5} \xi \vert W_0 \vert ^2}{4 \sqrt{2}
  g_s^{\frac{3}{2}} \sigma^{\frac{3}{2}}}. \ee The $\mc{O}(\frac{1}{g_s})$
bound on $\sigma$ comes from requiring  $\mc{V}_s > 1$ in order to
control the $\alpha'$ expansion. In general this regime is rather limited.
For very moderate values of $W_0$, equation (\ref{SigmaEquations}) has no solutions and there is \emph{no}
region of moduli space in which it is permissible to neglect $\alpha'$ corrections to
(\ref{OneModulusPotentials}).

For concreteness let us consider superpotentials generated by D3-brane
instantons and let us take\footnote{Properly this should be
  $e^{\mc{K}_{cs}/2}A$ to be K\"ahler covariant,
but we shall often drop the $e^{\mc{K}_{cs}/2}$.} $A=1$ and $g_s = \frac{1}{10}$.  Then we require $W_0 <
\mc{O}(10^{-75})$ to have any solutions at all of (\ref{SigmaEquations}) with $\mc{V}_s > 1$. This can be
improved somewhat by using gaugino condensation with very large gauge groups. Thus using $W_0 = 10^{-5}$, $g_s =
\frac{1}{10}$ and $N = 50$, $\sigma_{max} \sim 150$, which corresponds to $\mc{V}_s \sim 12$. However, as
generic values of $W_0$ are $\mc{O}(\sqrt{\frac{\chi}{24}}) \sim
\mc{O}(10)$ and $W_0^2$ is uniformly distributed \cite{hepth0404257}, $W_0
= 10^{-5}$ represents a tuning of one part in $10^{12}$, and even then the range of validity is rather limited.

Our conclusion is then that for generic values of $W_0$ there is no regime in which the perturbative corrections
to the K\"ahler potential can be neglected; the inclusion solely of non-perturbative corrections is
inconsistent. Furthermore, even for the small values of $W_0$ for which there does exist a regime in which
non-perturbative terms are the leading corrections, this is still only
true for a 
small range of moduli values and
in particular never holds at large volume.\footnote{We do not consider the special case $\chi(M) = 0$ , when the
$\alpha'^3$ corrections considered above vanish. We do not know the
  status of higher $\alpha'$ corrections in such models.}

\subsection{Systematics of the $\alpha'$ Expansion}
\label{WarpingSection}
The above arguments show that in KKLT-type flux compactifications, under almost all circumstances the neglect of
$\alpha'$ corrections is inconsistent. There are two conclusions to be drawn from this. First, to investigate
the structure of the potential at large volumes we must include the $\alpha'$ corrections. It is thus in
principle impossible to ask whether type IIB flux compactifications can realise the large extra dimensions
scenario without including these corrections. Secondly, the inclusion of the $\alpha'$ corrections is necessary
for the study of the type IIB flux landscape and its statistics. Models with
all moduli stabilised are a precondition for meaningful discussions of the landscape and its properties, and the
$\alpha'$ corrections are required to discuss moduli stabilisation across almost the entirety of K\"ahler moduli
space. While there exist examples for which the K\"ahler moduli can be stabilised purely by non-perturbative effects
\cite{hepth0503124}, 
such models are unlikely to be representative, not least because all such minima are
supersymmetric.

Fortunately, the inclusion of  $\alpha'$ corrections
does not require full knowledge of the string theory.
At small volumes, the $\alpha'$ corrections do appear democratically, and
it would be difficult to extract reliable results. However,
the $\alpha'$ expansion is at heart an expansion in inverse volume.
At very large volumes, the expansion parameter for $\alpha'$ effects is
$\frac{1}{\mc{V}}$, and a systematic inclusion of such effects
will give a controlled expansion.

Let us treat this more explicitly by studying the volume scaling of
the terms arising from dimensional reduction of the $\alpha'$-corrected type IIB supergravity action.
Of course this action is not known fully, but our arguments will only
depend on the general form of the terms rather than
the specific details of the tensor structure.
The supergravity action consists of bulk and localised terms and
is
\be
\label{FullAction}
S_{IIB} = S_{b,0} + \alpha'^3 S_{b,3} + \alpha'^4 S_{b,4}
+ \alpha'^5 S_{b,5} + \ldots +
S_{cs} + S_{l,0} + \alpha'^2 S_{l,2} + \ldots
\ee
The localised sources present are D3/D7 or O3/O7 planes. In string frame, we have \cite{Polchinski}
\bea
\label{iibaction}
S_{b,0} & = & \frac{1}{(2 \pi)^7 \alpha'^4} \int d^{10}x \sqrt{-g}
\left\{ e^{-2 \phi}[\mc{R} + 4(\nabla \phi)^2] - \frac{F_1^2}{2} -
  \frac{1}{2 \cdot 3!} G_3 \cdot \bar{G}_3 - \frac{\tilde{F}_5^2}{4
    \cdot 5!} \right\},
  \nonumber \\
S_{cs} & = & \frac{1}{4 i (2 \pi)^7 \alpha'^4} \int e^\phi C_4 \wedge
G_3 \wedge \bar{G}_3, \nonumber \\
S_{l,0} & = & \sum_{sources} \left( - \int d^{p+1} \xi \, T_p \, e^{-\phi} \sqrt{-g}
 + \mu_p \int C_{p+1} \right).
\eea
We may avoid the need to include
D3-branes by taking the fluxes to saturate the $C_4$ tadpole.
We shall work throughout in the F-theory orientifold limit, in which the
dilaton is constant: $\tau = \tau(y) = \tau_0$.

For flux compactifications with ISD 3-form fluxes, the metric and fluxes
take the form \cite{hepth0105097}
\bea
\label{metric5flux}
ds_{10}^2 & = & e^{2A(y)}\eta_{\mu \nu} dx^\mu dx^\nu + e^{-2A(y)}\tilde{g}_{mn}dy^m dy^n, \nonumber \\
\tilde{F}_5 & = & (1 + *) \left[ d\alpha \wedge dx^0 \wedge dx^1
  \wedge dx^2 \wedge dx^3 \right], \nonumber \\
F_3, H_3 & \in & H^3(M, \mbb{Z}), \eea where $\alpha = e^{4A}$ parametrises both the magnitude of the warping
and the size of the 5-form flux. Here $\tilde{g}$ is a Calabi-Yau metric; the flux back-reacts to render the
compact space only conformally Calabi-Yau. The warp-factor transforms non-trivially under internal rescalings,
with warping effects suppressed at large volume. Specifically, under $\mc{V} \to \lambda^6 \mc{V}$, where
$\lambda \gg 1$, $\alpha = 1 + \mc{O}(\frac{1}{\lambda^4}) + \ldots $.

We first consider contributions tree level in $\alpha'$.
The leading contribution to the 4-d scalar potential arises from the flux
term $\frac{1}{(2 \pi)^7 \alpha'^4} \int G_3 \cdot \bar{G}_3$,
which gives
\be
\label{Vflux}
V_{flux} \sim \mc{K}_{cs}^{a \bar{b}} \frac{D_a W D_{\bar{b}} W}{\mc{V}^2},
\ee
where the sum is over dilaton and complex structure moduli.
This term is positive semi-definite and vanishes at its minimum. The
volume scaling is understood as follows:
\be
V_{flux} \sim \overbrace{\mc{V}^{-2}}^{\textrm{conversion to Einstein frame}} \ti
\overbrace{\mc{V}}^{\textrm{internal integral}} \ti
\overbrace{\mc{V}^{-1}}^{G_3 \cdot \bar{G_3}} \sim \mc{V}^{-2}.
\ee

In the absence of warping, (\ref{Vflux}) is the only $\mc{O}(\alpha'^0)$ contribution to the potential energy,
as $\tilde{F}_5 = \mc{R} = 0$ and dilaton gradients vanish. However, there is also a warping contribution. At
large volume, $\alpha \sim 1 + \frac{1}{\mc{V}^{\frac{2}{3}}}$ and thus
 $V_{F_5} = \int d^6 x \sqrt{g} F_5^2  \sim \int d^6 x \sqrt{g} (\textrm{d} \alpha)^2$ contributes
\be
V_{F_5} \sim \overbrace{\mc{V}^{-2}}^{\textrm{conversion to Einstein frame}} \ti
\overbrace{\mc{V}}^{\textrm{internal integral}} \ti
\overbrace{\mc{V}^{-\frac{5}{3}}}^{F_5^2} \sim \mc{V}^{-8/3}.
\ee
As now $\mc{R} \neq 0$,
the Einstein-Hilbert term $\int_{Y_6} \sqrt{-g} \mc{R}$ is also
important and in fact contributes identically with $V_{F_5}$.
These terms may be related to the tree-level flux term and serve as an
additional prefactor. The net result is \cite{hepth0208123}
$$
V_{0,unwarped} = \frac{1}{2 \kappa_{10}^2 \textrm{Im } \tau} \int_M G_3^{+} \wedge *_6 \bar{G}_3^{+}
\to V_{0, warped} = \frac{1}{2 \kappa_{10}^2 \textrm{Im } \tau} \int_M e^{4A} G_3^{+} \wedge *_6 \bar{G}_3^{+}.
$$
It is important that this potential remains no-scale,
with no potential generated for the  K\"ahler moduli.

The bulk effective action receives higher-derivative corrections
starting at $\mc{O}(\alpha'^3)$, which is also the order at which
string loop corrections first appear;
the tree level action is already $SL(2, \mbb{Z})$ invariant and
receives no $g_s$ corrections.
The discussion of loop corrections to $S_b$ is thus subsumed into the
discussion of higher-derivative corrections.

The bosonic fields are the metric, dilaton-axion,
3-form field strength $G_3$ and self-dual five form field strength
$F_5$. While its precise form is unknown, $S_{b,3}$ is expected to include all combinations of
these consistent with the required dimensionality.
At $\mc{O}(\alpha'^3)$, the bosonic action takes the schematic form
\bea
\label{Sb3}
S_{b,3} & \sim & \frac{\alpha'^3}{\alpha'^4} \int d^{10}x \sqrt{-g}
\Big[ \Big( \mc{R}^4 + \mc{R}^3 \left( G_3 G_3 + G_3 \cdot \bar{G}_3 + G_3
\bar{G}_3 + F_5^2 + \partial \tau \cdot \partial \tau + \nabla^2 \tau
\right) \nonumber \\
& & + \mc{R}^2((DG_3)^2 + (DF_5)^2 + G^4 + \ldots ) 
+ \mc{R}(G_3^6 + \ldots ) + (G_3^8 + \ldots)\Big) \Big]. \nonumber
\eea
Terms linear in the
fluxes (e.g. $\mc{R}^3 DG_3$) are forbidden as the action must be
invariant under world-sheet parity.
The tensor structure and modular behaviour of the majority of these terms is unknown. A notable exception is
the $\mc{R}^4$ term, the coefficient of which is known exactly to be
an Eisenstein series in the dilaton \cite{hepth9808061}.
However, our interest is in the volume scaling, which can be extracted
on merely dimensional grounds.

Let us first consider terms independent of warping.
The $\mc{R}^3 (G^2 + c.c.)$ and $\mc{R}^3 G \bar{G}$
terms are most easily understood. These are similar to the
$\mc{O}(\alpha'^0)$ $G_3 \bar{G_3}$ term but with three extra powers of curvature.
Then
\be
V_{\mc{R}^3 G^2} \sim \overbrace{\mc{V}^{-2}}^{\textrm{ Einstein frame}} \ti
\overbrace{\mc{V}}^{\textrm{internal integral}} \ti \overbrace{\mc{V}^{-1}}^{\mc{R}^3} \ti
\overbrace{\mc{V}^{-1}}^{G_3^2} \sim \mc{V}^{-3}.
\ee
The same argument tells us that a similar scaling applies for
$\mc{R}^2 (DG_3)^2$ terms, whereas
$\mc{R}^2 G^4$ terms contribute
as $\sim \frac{1}{V^{11/3}}$, $\mc{R} G^6$ terms as $\sim \frac{1}{V^{13/3}}$ and $G^8$ terms as
$\sim \frac{1}{V^5}$.
In the absence of warping, the $\mc{R}^4$ term does not contribute to the potential
energy; integrated over a Calabi-Yau, it vanishes.
This geometric result can be understood macroscopically;
were this not to vanish, it would generate a potential for the
volume even in flux-less $\mc{N} = 2$ IIB compactifications. However,
it is known there that tree level moduli remain moduli to all orders in
the $\alpha'$ and $g_s$ expansions, and thus higher derivative
terms must not be able to generate a
potential for them. This term contributes indirectly by modifying
the dilaton equations of motion at $\mc{O}(\alpha'^3)$; we will return to this later.

There are also higher-derivative terms dependent on the warp factor.
Examples are
$$
\frac{\alpha'^3}{\alpha'^4} \int \sqrt{g} \mc{R}^4, \quad \quad
\frac{\alpha'^3}{\alpha'^4} \int \sqrt{g} \mc{R}^3 F_5^2, \quad \quad
\frac{\alpha'^3}{\alpha'^4} \int \sqrt{g}
\mc{R}^2 (DF_5)^2.
$$
These are similar to the corresponding tree-level terms but
with three extra powers of curvature.
As $\mc{R}^3 \sim \frac{1}{\mc{V}}$, the contribution of such terms is
no larger than $\mc{O}(\frac{1}{\mc{V}^{11/3}})$.

There are also potential contributions from
internal dilaton gradients.
As we have worked in the orientifold limit of F-theory, we have
thus far regarded such terms as vanishing.
However, this is not quite true.
In the presence of higher derivative corrections,
a constant dilaton no longer solves the
equations of motion. Instead, we have
\be
\label{newdilaton}
\phi(y) = \phi_0 + \frac{\zeta(3)}{16} Q(y).
\ee
An explicit expression
for $Q(y)$ may be found in \cite{hepth0204254},
but for our purposes it is sufficient to note that $Q \sim \mc{R}^3 \sim
\frac{1}{\mc{V}}$. It is then easy to see that terms such as
$$
\int (\partial \tau \cdot \partial \bar{\tau}) \mc{R}^2 G^2 \quad
\quad \textrm{ or } \quad \quad
\int (\nabla^2 \phi) (DF_5)^2 \mc{R}
$$
are suppressed compared to the terms considered above.
Note that dilaton-curvature terms such as
$$
\int (\nabla^2 \phi) \mc{R}^3
$$
will not contribute to the potential energy; these exist in $\mc{N} =
2$ Calabi-Yau compactifications
and so must vanish either directly or by cancellation. Similar comments apply for the fact that, even
without warping, the internal space ceases to be Calabi-Yau at $\mc{O}(\alpha'^3)$.

There is one further effect associated with (\ref{newdilaton}). As the
dilaton is no longer constant, under dimensional reduction
the four-dimensional Einstein-Hilbert term is renormalised.
Rescaling this to canonical form introduces a term $\frac{V_{tree}}{\mc{V}}$
of $\mc{O}(\frac{1}{\mc{V}^3})$.
However, as $V_{tree}$ is no-scale this correction does not break the
no-scale structure and in particular vanishes at the minimum.

String loop corrections first appear at $\mc{O}(\alpha'^3)$ and are thus subsumed into the above analysis. While
it may be difficult to derive anything explicitly, we may conjecture their effect in the large volume limit. The
corrections to the K\"ahler potential arise from dimensional reduction of the 10-dimensional $\mc{R}^4$ term. In
Einstein frame, the string tree-level  $\alpha'^3$ corrected K\"ahler potential derived in \cite{hepth0204254}
takes the form \be \label{Kalpha'} \K = \mc{K}_{cs} -2 \ln \left(\mc{V} + \frac{\xi}{2 e^{3 \phi/2}}\right). \ee
Here $\xi  = - \frac{\chi(M) \zeta(3)}{2(2 \pi)^3}$. $\zeta(3)$ is distinctive as the tree-level coefficient of
the ten-dimensional $\mc{R}^4$ term, the coefficient of which is known exactly to be \be
f^{(0,0)}_{\frac{3}{2}}(\tau, \bar{\tau}) = \sum_{(m,n) \neq (0,0)} \frac{e^{-\frac{3 \phi}{2}}}{\vert m + n \tau \vert^3}.
\ee This has the expansion \be f^{(0,0)}_{\frac{3}{2}}(\tau, \bar{\tau}) = \frac{2 \zeta(3)}{e^{\frac{3
      \phi}{2}}} + \frac{2 \pi^2}{3} e^{\frac{\phi}{2}} + \textrm{ instanton
  terms }.
\ee
Therefore, to incorporate $\mc{O}(\alpha'^3)$ string loop corrections
to $S_{IIB}$,
the natural conjecture is that we should modify (\ref{Kalpha'}) to
\be
\K = \K_{cs} - 2 \ln \left(\mc{V} - \frac{\chi(M)}{8(2 \pi)^3}
f^{(0,0)}_{\frac{3}{2}}(\tau, \bar{\tau}) \right).
\ee

Let us finally mention further higher derivative corrections at
$\mc{O}(\alpha'^4)$ and above. At large volume these are all
subleading, with any terms generated being
subdominant to the $\frac{1}{\mc{V}^3}$ terms present at
$\mc{O}(\alpha'^3)$. For example, an $\mc{O}(\alpha'^4)$ term
$G^2 \mc{R}^4$ would give a $\mc{V}^{-\frac{10}{3}}$ contribution to
the potential.
There are other terms that would naively give a $\frac{1}{\mc{V}^3}$
contribution, such as a possible $\mc{O}(\alpha'^6)$ term $\mc{R}^6$.
However, on a Calabi-Yau such a term vanishes, either
explicitly or through cancellation. This is for the reasons discussed above:
terms present even in pure $\mc{N} = 2$
Calabi-Yau compactification cannot generate potentials for the moduli.

Our conclusion is therefore that the leading $\alpha'$ corrections
breaking the no-scale structure appear at $\mc{O}(\frac{1}{\mc{V}^3})$, coming from
$\mc{R}^3 G^2$ and $\mc{R}^2 (DG)^2$ terms in ten dimensions. This is
consistent with the result of \cite{hepth0204254}, where the
corresponding correction (\ref{Kalpha'}) to the K\"ahler potential was computed
and the resulting scalar potential interpreted as descending from such
ten-dimensional terms. Although the scaling argument used above cannot reproduce the
coefficient of the correction, it makes it easier to see that other
terms are subleading; in particular, the effects associated
with non-vanishing $F_5$ (which were not considered in \cite{hepth0204254}) do not compete.

Let us now consider higher derivative
corrections to localised sources.
The D3-brane action is
\be
S_{D3} = -T_3 \int d^4x \sqrt{-g} e^{-\phi} + \mu_3 \int C_4.
\ee
As D3-branes are space-filling, any higher derivative corrections involve space-time
curvature and so cannot give a potential energy. Furthermore, it is
known that the D3-brane solution is unaltered by the $\alpha'^3$ flux
and curvature terms \cite{hepth0308061}.
The $\alpha'$ corrections on D7-branes do give potential energy contributions;
fortunately in that case such effects were already included in the
analysis of \cite{hepth0105097}. We briefly review the relevant results.
The leading $\alpha'$ correction to
the D7-brane Chern-Simons action is (we do not turn on internal D7-brane
fluxes)
\be
S_{D7, \alpha'^2} = \frac{\mu_7}{96}(2 \pi \alpha')^2 \int_{R^4 \ti \Sigma} C_4 \wedge \textrm{Tr}( \mc{R}_2 \wedge \mc{R}_2),
\ee
whereas the leading correction to the wrapped D7-brane tension arises from a term
\be
\frac{-\mu_7}{96}(2 \pi \alpha')^2 \int_{R^45 \ti \Sigma} \sqrt{-g} \textrm{Tr}(\mc{R}_2 \wedge *\mc{R}_2).
\ee
These contribute effective D3-brane charge and tension. In F-theory,
the D3-brane charge from the wrapped D7-branes is
\be
\label{FD3charge}
Q_{D3} = -\frac{\chi(X)}{24},
\ee
where $X$ is the Calabi-Yau fourfold. As the D7 branes are BPS, (\ref{FD3charge})
also gives the resulting effective D3 tension.

A similar argument also shows that
higher $\alpha'$ corrections to the D7-brane action need not be
considered.
As the branes are BPS, $\alpha'$ corrections to the tension are related
to $\alpha'$ corrections to the charges. However, these
involve spacetime curvature, and so the corresponding correction to
the DBI action cannot give rise to a 4d potential energy.

We have focused above on $\alpha'$ corrections from bulk and localised
sources, and $g_s$ corrections from the bulk.
We should finally consider string loop corrections arising from the open string sector.
Less is known about these effects and it would be very interesting to study
them further (see \cite{bhk}). However, there is a sense in which such
effects could only strengthen our results. We have already shown that the use solely
of non-perturbative effects to describe the moduli potential is
generically inconsistent due to the competing bulk $\alpha'^3$
effects; extra corrections, if they exist, would further enhance the
importance of perturbative effects.

\section{Spectrum in the Moduli Sector}
\linespread{1.2}

\subsection{A working model}
\label{WorkingModel}

We have shown above both that $\alpha'$ corrections must be
included to study K\"ahler moduli stabilisation, and that at
large volume we only need include the leading corrections of
\cite{hepth0204254}. In \cite{hepth0502058} a general analysis
of the resulting potential was carried out. It was shown that there
exists a special decompactification direction in moduli space along which the potential
vanishes from below, with an associated minimum at
exponentially large volumes. This behaviour was exhibited
explicitly in a particular model, flux compactificatons on the
orientifold of $\mbb{P}^4_{[1,1,1,6,9]}$.

The general argument for this behaviour will be reviewed
and given a geometric interpretation in section \ref{GenSection};
however, we shall first extend our study of the concrete example. In particular,
we shall calculate the spectrum of scales and particle masses,
as well as the soft supersymmetry breaking
terms.

$\mbb{P}^4_{[1,1,1,6,9]}$ has $h^{1,1} = 2$ and $h^{2,1} = 272$ and
its K\"ahler sector  has been studied in \cite{hepth0404257}.
In terms of the K\"ahler moduli, which following the
notation of \cite{hepth0404257} we denote by $T_4$ and $T_5$, the volume may be written as
\be
\label{ModelVol}
{\cal{V}} = \frac{1}{9\sqrt{2}} \left(\tau_5^{\frac{3}{2}} - \tau_4^{\frac{3}{2}}\right),
\ee
where $T_4 = \tau_4 + i b_4$ and $T_5 = \tau_5 + i b_5$.
We fix the dilaton and complex structure moduli using 3-form
fluxes. The periods are known \cite{hepth9403187}
and this can be done straightforwardly as in \cite{hepth0312104, hepth0404243, hepth0409215}.
We then integrate these out and concentrate on the K\"ahler moduli.
It was shown in \cite{hepth0404257} that the divisors $\mc{D}_i$ corresponding to $\tau_4$ and $\tau_5$
have $\chi(\mc{D}) = 1$ and thus non-perturbative superpotentials will be generated
for these moduli \cite{wittenw}.

The condition $\chi(D) = 1$ for $D$ a vertical divisor in the absence
of flux is a necessary but not a sufficient
condition for divisors to contribute to a superpotential. In the
presence of fluxes, instanton zero modes can be lifted and
non-perturbative superpotentials generated under more general
circumstances. The exact necessary conditions are not yet known, but
the trend of recent results \cite{hepth0407130, hepth0503072,
  hepth0501081, hepth0503125, hepth0503138, hepth0504058} 
is that the condition $\chi(D) =
1$ will be substantially relaxed. However, in the model we study the
non-perturbative superpotentials exist even under the most
conservative requirement.

After integrating out dilaton and complex structure moduli, the resulting superpotential is
\be
\label{ModelSPot}
W = W_0 + A_4 e^{-\frac{a_4}{g_s} T_4} + A_5 e^{-\frac{a_5}{g_s} T_5}.
\ee
There is a numerical prefactor of $g_s^2 M_P^3$ in (\ref{ModelSPot})
which will be important for the discussion of moduli masses in the
next section.
$W_0$ is determined by the fluxes; while tunable in principle, we impose no requirement that
it be small.
The K\"ahler potential is
\be
\label{ModelKPot}
\K = \K_{cs} - 2 \ln \left({\cal{V}} + \frac{\xi}{2}\right),
\ee
where the volume ${\cal{V}}$ is given by (\ref{ModelVol}). Here $g_s = \langle e^\phi \rangle$ and Einstein
frame has been defined by $g_{\mu \nu, s} = e^{(\phi - \phi_0)/2} g_{\mu \nu, E}$ (note that
this convention differs from our usage in section \ref{typeiibsection}).

\begin{figure}[ht]
\linespread{0.2}
\begin{center}
\epsfxsize=0.75\hsize \epsfbox{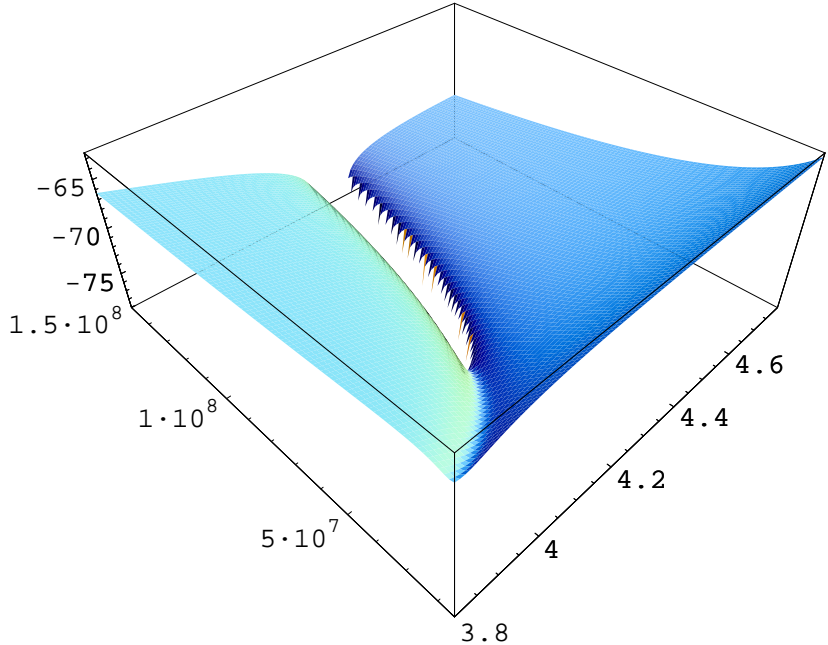}
\end{center}
\caption{ $\ln (V)$ for ${\mathbb{P}}^4_{[1,1,1,6,9]}$ in the large volume
  limit, as a function of the divisors $\tau_4$ and $\tau_5$. The void
  channel corresponds to the region
  where $V$ becomes negative and $\ln (V)$ undefined. As $V \to 0$ at
  infinite volume, this immediately shows that a large-volume minimum
  must exist.}
\label{LargeVolumeLimit}
\end{figure}

Given (\ref{ModelSPot}) and (\ref{ModelKPot}), the scalar potential (\ref{FScalarPot})
may be computed directly. In the limit where $\tau_5 \gg \tau_4 > 1$, this takes the form
\be
\label{LargeTau5Pot}
V = \frac{\lambda \sqrt{\tau_4} (a_4 A_4)^2 e^{-2 \frac{a_4
    \tau_4}{g_s}}}{\cal{V}} - 
\frac{\mu W_0 (a_4 A_4) \tau_4 e^{- \frac{a_4 \tau_4}{g_s}}}{{\cal{V}}^2}
+ \frac{\nu \xi W_0^2}{{\cal{V}}^3},
\ee
where $\lambda$, $\mu$ and $\nu$ are model-dependent numerical constants. The axion partner of $\tau_4$ will
adjust to render the middle term of (\ref{LargeTau5Pot}) negative; this adjustment reduces the potential, 
and will always occur.
At large volumes we would naively expect the potential to be uniformly positive,
due to the perturbative $\alpha'^3$ term, but if $\tau_4$ is logarithimically small
compared to $\tau_5$ this need not be true. Consider the decompactification limit
$$
\tau_5 \to \infty \quad \textrm{ and } \quad a_4 A_4 e^{- \frac{a_4 \tau_4}{g_s}} = \frac{W_0}{\mc{V}}.
$$
In this limit the potential (\ref{LargeTau5Pot}) becomes
\be
V = \frac{W_0^2}{{\cal{V}}^3}\left(\lambda \sqrt{\ln(\mc{V})} - \mu \ln \mc{V} + \xi \nu\right)
\ee
and thus approaches zero from below. Negativity of the potential requires that $\ln \mc{V}$ be large, and
so associated with this behaviour there exists a
minimum at exponentially large volumes. This is illustrated in Figure \ref{LargeVolumeLimit}.
It is possible to solve
$$
\frac{\partial V}{\partial \tau_4} = \frac{\partial V}{\partial \tau_5}  = 0
$$
analytically, to obtain
\begin{equation}
\label{VolumeAtMin}
\tau_4 \propto \xi^{\frac{2}{3}} \quad \textrm{ and } \quad \langle
    {\cal{V}} \rangle
\propto W_0 e^{\frac{a_4 \tau_4}{g_s}}.
\end{equation}
It is necessary that this remains a minimum of the full potential including the dilaton and
complex structure moduli. These are flux-stabilised and enter the potential as $\frac{DW \cdot DW}{\mc{V}^2}$; at the
minimum this term vanishes, but is otherwise $\mc{O}(\frac{1}{\mc{V}^{2}})$.
However, the vacuum energy at the minimum is $-\mc{O}(\frac{1}{{\cal{V}}^3})$. This ensures that the minimum is
a real minimum of the full potential: movement of the complex
structure moduli away from their stabilised values contributes a term $ + \mc{O}(\frac{1}{\mc{V}^2})$
to the potential, but the negative terms cannot compete with this, and so this can only increase the potential.
Therefore the minimum remains a minimum of the full potential
\footnote{In generic Calabi-Yau orientifold compactifications,
  in addition to the dilaton, complex structure and K\" ahler
  moduli, the resulting four dimensional low energy theory also has $h_-^{1,1}$
  moduli coming from the reduction of the Type
  IIB 2-forms $B_2$ and $C_2.$ We will assume throughout that these
  are not present; in particular, this is true for the
  ${\mathbb{P}}^4_{[1,1,1,6,9]}$ model we discuss below.}.

\subsection{Scales and Moduli Masses}
\label{MassesSection}
We set $\hbar = c = 1$ but will otherwise be pedantic on frames and factors of $2 \pi$ and $\alpha'$.
Our basic length will be $l_s = 2 \pi \sqrt{\alpha'}$ and our basic mass $m_s = \frac{1}{l_s}$.
These represent the only dimensionful scales, and unless specified
otherwise volumes are measured in units of $l_s$. We will furthermore require that
at the minimum the 4-dimensional metric is the metric the string worldsheet couples to.

Stringy excitations then have
\be
m_S^2 = \frac{n}{\alpha'} \Rightarrow m_S \sim 2 \pi m_s.
\ee
To estimate Kaluza-Klein masses we first recall  toroidal
compactifications. A stringy ground state of Kaluza-Klein and winding integers
$n$ and $w$ has mass
\be
\label{KKmasses}
m_{KK}^2 = \frac{n^2}{R^2} + \frac{w^2 R^2}{\alpha'^2},
\ee
where $R$ is the dimensionful Kaluza-Klein radius.
Strictly (\ref{KKmasses}) only holds for toroidal compactifications, but it should
suffice to estimate the relevant mass scale.
If we write $R = R_s l_s$ and assume $R_s \gg 1$, we have
\be
\label{NewKKmasses}
m_{KK} \sim \frac{m_s}{R_s} \quad \textrm{ and } \quad m_W \sim (2 \pi)^2 R_s m_s.
\ee
It is conceivable that the geometry of the internal space is elongated such
that the Kaluza-Klein radius $R_s$ is
uncorrelated with the overall volume, and we would then have
KK masses of order $m_{KK}^4\sim 1/\tau_i$ for
the different cycles.
However, in absence of evidence to the contrary we assume
the simplest scenario in which $(2 \pi R_s)^6 = V_s$. 
Then
\be
\label{KKMass}
m_{KK} \sim \frac{2 \pi m_s}{{\cal V}_s^{\frac{1}{6}}}. 
\ee
Here $m_{KK}$ refers only to the lightest KK mode, as in our
situation the overall volume is large but there are
relatively small internal cycles. Therefore, while there may be many
KK modes, there is a hierarchy  with the others being naturally
heavier than the scale
of (\ref{KKMass}).

We next want to determine the masses of the complex structure and K\"ahler moduli.
This requires an analytic expression for the potential
in terms of canonically normalised fields.
To obtain the potential, we will work in the framework of $\mc{N} =
1$ supergravity; the validity of this effective field theory approach is discussed in
section \ref{EFT}. The dimensional reduction of the 10-dimensional action into this
framework is carried out in more detail in 
appendix A; here we shall just state results.

An $\mc{N} = 1$ supergravity is completely specified by a K\"ahler potential, superpotential and gauge kinetic function.
Neglecting the gauge sector to focus on moduli dynamics, the action is
\be
S_{\mc{N}=1} = \int d^4 x \sqrt{-G} \left[ \frac{M_P^2}{2} \mc{R} - \K_{,i\bar{j}} D_\mu \phi^{i} D^\mu \bar{\phi}^j
- V(\phi, \bar{\phi}) \right],
\ee
where
\be
\label{ScalarPotential}
V(\phi, \bar{\phi}) = e^{\K/M_P^2} \left(\K^{i \bar{j}} D_i \hat{W} D_{\bar{j}} \bar{\hat{W}} - \frac{3}{M_P^2} \hat{W}
\bar{\hat{W}} \right) + \textrm{ D-terms}.
\ee
$\K$ is the K\"ahler potential, which has mass dimension 2, and $\hat{W}$ the superpotential, with mass dimension 3.
$M_P$ is the reduced Planck mass $M_P = \frac{1}{(8 \pi G)^\half} =
2.4 \ti 10^{18} \textrm{GeV}$ and the Planck and string scales are related by
\be
M_P^2 = \frac{4 \pi \mc{V}_s^0}{g_s^2 l_s^2} \quad \textrm{ or } \quad
m_s = \frac{g_s}{\sqrt{4 \pi \mc{V}_s^0}} M_P.
\ee
Here $\mc{V}_s^0 = \langle \mc{V}_s \rangle$ is the string-frame
volume at the minimum.
As shown in the appendix, including the $\alpha'$ and non-perturbative corrections, the
K\"ahler and superpotentials are
\bea
\label{Potentials}
\frac{\mc{K}}{M_P^2} & = & - 2 \ln \left(\mc{V}_s + \frac{ \xi g_s^{\frac{3}{2}}}{2 e^{\frac{3 \phi}{2}}} \right)
- \ln(-i(\tau - \bar{\tau})) - \ln \left(-i \int_{CY} \Omega \wedge \bar{\Omega}\right), \nonumber \\
\hat{W}  & = & \frac{g_s^{\frac{3}{2}} M_P^3}{\sqrt{4 \pi} l_s^2} \left( \int_{CY}
 G_3 \wedge \Omega + \sum A_i e^{\frac{- 2 \pi}{g_s} T_i} \right)
\equiv \frac{g_s^{\frac{3}{2}} M_P^3}{\sqrt{4 \pi}} W.
\eea
Here $\xi = -\frac{\zeta(3) \chi(M)}{2 (2 \pi)^3}$ and
$T_i = \tau_i + i b_i$, where $\tau_i = \int d^4 x
\sqrt{\tilde{g}}$ is a  4-cycle volume and $b_i$ its axionic
partner arising from the RR 4-form;
these are good K\"ahler coordinates for IIB orientifold compactifications.
The factors of $g_s$
are unconventional and arise from our definition of the 10-dimensional Einstein
frame (\ref{Einsteinframe}) ($g_{\mu \nu, s} = e^{(\phi - \phi_0)/2} g_{\mu \nu, E }$).
The advantage of this definition is that the resulting volumes are
a true measure of `largeness', and the stringy non-perturbative nature of the
$e^{-\frac{a_i}{g_s} T_i}$ term is manifest.

To calculate scalar masses, we must express the potential (\ref{ScalarPotential}) in terms of
canonically normalised fields.
The K\"ahler metric for the complex
structure moduli is given by
\be
\label{CSmetric}
G_{i \bar{j}} = \partial_i \partial_{\bar{j}} \mc{K}_{cs} = \partial_i
\partial_{\bar{j}} \ln \left( - i \int_{CY} \Omega \wedge \bar{\Omega} \right).
\ee
In the no-scale approximation, which holds to leading order, the potential for the complex structure moduli is
\be
V = \frac{g_s^4 M_P^4}{8 \pi (\mc{V}_s^0)^2} \int d^4 x \sqrt{-g_E} e^{\mc{K}_{cs}} \left[ G^{a \bar{b}}
 D_a W D_{\bar{b}} \bar{W}  \right],
\ee
where the sum runs over complex structure moduli only. 
The inverse of (\ref{CSmetric}) is hard to make explicit, as to do so
would require knowledge of all the Calabi-Yau
periods.
However, as (\ref{CSmetric}) is independent of dilaton and K\"ahler
moduli, this process will not introduce
extra factors of $\mc{V}_s$ or $g_s$. Thus if we assume numerical factors
to be $\mc{O}(1)$, we find
\be
m_{cs}^2 = \mc{O}(1) \frac{g_s^4 N^2 M_P^2}{4 \pi ({\cal V}_s^0)^2},
\ee
where $N \sim \mc{O}(\sqrt{\frac{\chi}{24}})$ is a measure of the typical number of flux quanta
and arises from the $D_a W$ terms.
We therefore have
\be
m_{cs} = \mc{O}(1) \frac{g_s N m_s}{\sqrt{\mc{V}_s^0}}.
\ee
As emphasised in \cite{hepth0309170}, one requires a clear separation between
Kaluza-Klein and complex structure masses to trust the supergravity analysis.
We have
\be
\label{csKK}
\frac{m_{cs}}{m_{KK}} \sim \frac{g_s N}{2 \pi ({\cal V}_s^0)^{\frac{1}{3}}}.
\ee
At large volumes, this ratio is much less than one, which is reassuring.

For the concrete $\mbb{P}^4_{[1,1,1,6,9]}$ example,
the K\"ahler moduli may be treated more explicitly.
It is nonetheless hard to normalise the fields canonically across the entirety of
moduli space. However, to compute the spectrum we only need
normalise the moduli at the physical minimum.
It turns out (see appendix B) that the appropriately normalised fields are
\bea
\label{CanonicalFields}
\tau_5^{c} = \sqrt{\frac{3}{2}} \frac{\tau_5}{\tau_5^0} M_P, & &
b_5^{c} = \sqrt{\frac{3}{2}} \frac{b_5}{\tau_5^0} M_P, \nonumber \\
\tau_4^{c} = \sqrt{\frac{3}{4}} \frac{\tau_4}{(\tau_5^0)^{\frac{3}{4}}
  (\tau_4^0)^{\frac{1}{4}}} M_P, & &
b_4^{c} = \sqrt{\frac{3}{4}} \frac{b_4}{(\tau_5^0)^{\frac{3}{4}} (\tau_4^0)^{\frac{1}{4}}} M_P.
\eea
Here $\tau_5^0 = \langle \tau_5 \rangle$, etc. The bosonic mass matrix follows
by taking the second derivatives of the scalar potential with respect to $\tau_i^c$ and $b_i^c$.
In the vicinity of the large volume minimum, the scalar potential takes the form
\be
\label{Vgeneral}
V = g_s^4 M_P^4 \left(\frac{\lambda' \sqrt{\tau_4} e^{-2 \frac{a_4 \tau_4}{g_s}}}{\tau_5^{\frac{3}{2}}}
+ \frac{\mu'}{\tau_5^3} \tau_4 e^{- \frac{a_4 \tau_4}{g_s}}\cos( \frac{a_4 b_4}{g_s}) + \frac{\nu'}{\tau_5^{\frac{9}{2}}} \right).
\ee
As discussed in \cite{hepth0502058} we have
$$
\lambda' \sim \frac{a_4^2 \vert A_4 \vert^2}{g_s^2}, \quad \mu' \sim \frac{a_4 \vert A_4 W_0 \vert}{g_s},
\textrm{ and } \nu' \sim \xi \vert W_0 \vert^2.
$$
The $b_5$ axion appears in terms suppressed by $e^{-a_5 \tau_5}$ and has not been
written explicitly. As $\tau_5^0 \gg 1$, it follows that this field is
essentially massless.
In terms of the canonical fields, the scalar potential
(\ref{Vgeneral}) becomes
\be
V = \frac{\lambda \sqrt{\tau_4^{c} \beta}
e^{-2 a_4 \beta
  \tau_4^{c}}}{(\tau_5^0)^{\frac{3}{2}}
(\tau_5^{c})^{\frac{3}{2}}}
 + \frac{\mu \tau_4^{c} \beta
e^{-a_4 \beta \tau_4^{c}}
\cos(a_4 \beta b_4)}{(\tau_5^0)^3 (\tau_5^{c})^3} +
\frac{\nu}{(\tau_5^0)^{\frac{9}{2}}(\tau_5^{c})^{\frac{9}{2}}},
\ee
where $\tau_4 = \beta g_s \tau_4^c$.
The mass matrix is
\be
\textrm{d}^2 V =
\left( \begin{array}{ccc} \frac{\partial^2 V}{\partial \tau_5^{c} \partial \tau_5^{c}} & \frac{\partial^2 V}{\partial
    \tau_5^{c} \partial \tau_4^{c}} & \frac{\partial^2 V}{\partial
    \tau_5^{c} \partial b_4^{c}} \\   \frac{\partial^2 V}{\partial
    \tau_4^{c} \partial \tau_5^{c}} &  \frac{\partial^2 V}{\partial
    \tau_4^{c} \partial \tau_4^{c}} & \frac{\partial^2 V}{\partial
    \tau_4^{c} \partial b_4^{c}} \\ \frac{\partial^2 V}{\partial
    b_4^{c} \partial \tau_5^{c}}& \frac{\partial^2 V}{\partial
    b_4^{c} \partial \tau_4^{c}} & \frac{\partial^2 V}{\partial
    b_4^{c} \partial b_4^{c}} \end{array}
\right).
\ee
At the minimum this
mass matrix takes the schematic form
$$
\frac{M_P^2 g_s^4}{({\cal V}_s^0)^2} \left( \begin{array}{ccc}
  \frac{a}{{\cal V}_s^0} & \frac{b}{({\cal V}_s^0)^{\half} g_s} & 0 \\
\frac{b}{({\cal V}_s^0)^{\half} g_s} & \frac{c}{g_s^2} & 0 \\ 0 & 0 & d \end{array} \right).
$$
Cross terms involving the axion decouple and we obtain
\bea
\label{Masses}
m_{\tau_5^{c}} = \mc{O}(1) \frac{g_s^2 W_0}{\sqrt{4 \pi}
  ({\cal V}_s^0)^{\frac{3}{2}}} M_P, & &  m_{b_5^{c}} \sim \exp (-\tau_5^0),
\nonumber \\
m_{\tau_4^{c}} = \mc{O}(1) \frac{a_4 g_s W_0}{\sqrt{4 \pi} {\cal V}_s^0}
M_P, & &  m_{b_4^{c}} = \mc{O}(1)
\frac{a_4 g_s W_0}{\sqrt{4 \pi}{\cal V}_s^0} M_P.
\eea
The division of scales between the large modulus ($\tau_5$) and the small
modulus ($\tau_4$) will turn out to be general (see section \ref{GenSection}).
The $\mc{O}(1)$ factors depend on the detailed geometry of the
particular Calabi-Yau and are therefore not written explicitly, although
given the K\"ahler potential their numerical
computation is straightforward.

The $\tau_4$ and $b_4$ moduli have masses similar to the dilaton and
complex structure moduli.
We may then worry that our above treatment was inconsistent, as
we first integrated out complex structure moduli
and only then considered K\"ahler moduli. However, as discussed in
section \ref{WorkingModel},
from the form of the full potential we see
that the solution derived by first integrating out
the complex structure moduli
remains a minimum of the full potential.

\subsection{Fermion Masses}

The fermions divide into the gravitino and the fermionic partners of
the chiral superfields. The gravitino mass is given by
\be
m_{\frac{3}{2}} = e^{\K/2} \vert \hat{W} \vert = \frac{g_s^2
  e^{\frac{\K_{cs}}{2}} \vert W_0 \vert}{{\cal V}_s^0 \sqrt{4 \pi}} M_P.
\ee
with $\K$ and $\hat{W}$ given by (\ref{Potentials}).

We use expressions appropriate for a Minkowski minimum and so assume
we have included lifting terms.
The mass matrix for the other fermions is then $[M_\psi]_{ij}
\bar{\psi}_{Li} \psi_{Lj}$, where
$[M_\psi]_{ij} = \sum_{n=1}^4 [M_\psi^n]_{ij}$, with
\bea
\label{FermionMasses}
\left[ M_{\psi}^{1} \right]_{ij} & = & -e^{\K/2} \vert \hat{W} \vert \left\{ \K_{ij} + \frac{1}{3}
  \K_i \K_j \right\} M_P, \nonumber \\
\left[ M_{\psi}^{2} \right]_{ij} & = & -e^{\K/2} \vert \hat{W} \vert \left\{ \frac{\K_i \hat{W}_j +
     \K_j \hat{W}_i}{3 \hat{W}} - 2 \frac{\hat{W}_i \hat{W}_j}{3
  \hat{W}^2} \right\} M_P, \nonumber \\
\left[ M_{\psi}^{3} \right]_{ij} & = & - e^{\K/2}
  \sqrt{\frac{\bar{\hat{W}}}{\hat{W}}} \hat{W}_{ij} M_P, \nonumber \\
\left[ M_{\psi}^{4} \right]_{ij} & = & e^{\mc{G}/2} \mc{G}_l
  (\mc{G}^{-1})_k^l \mc{G}_{ij}^k M_P,
\eea
where $\mc{G} = \K + \ln(\hat{W}) + \ln(\bar{\hat{W}})$, $\hat{W}_i = \partial_i \hat{W}$, $\K_i
  = \partial_i \K$, $\mc{G}^l = \partial_{\bar{l}} \mc{G}$,
  etc. Here derivatives are with respect to the canonically normalised
  fields (\ref{CanonicalFields}).

There is one massless fermion corresponding to the goldstino,
which is eaten by the gravitino. Essentially this is
$\tilde{\tau}_5$, although there is some small mixing with $\tilde{\tau}_4$.
The mass of $\tilde{\tau}_4$ can be calculated from
(\ref{FermionMasses}); we find
\be
m_{\tilde{\tau_4}} \approx \frac{g_s^2 a_4 W_0}{{\cal V}_s} M_P \approx m_{\tau_4} \approx m_{3/2}.
\ee
As with the bosonic spectrum, it is hard to obtain explicit expressions for modulino masses
for the complex structure moduli. However, there is no explicit volume dependence in $\K_{cs}$ or $W_{flux}$,
and so the volume dependence of $m_{\tilde{\phi}}$ is determined by the $e^{\frac{\K}{2}}$ terms. Therefore
\be
m_{\tilde{\phi}} \sim \frac{g_s^2 W_0}{{\cal V}_s} M_P, \qquad m_{\tilde{\tau}}
\sim \frac{g_s^2 W_0}{\mc{V}} M_P.
\ee
and modulino masses have a scale 
set by the gravitino mass. Thus, as expected, $m_{3/2}$ determines the scale of Bose-Fermi
splitting. As at large volume $m_{3/2} << m_s, m_{KK}$, the moduli and
modulino physics should decouple from that associated with stringy or
Kaluza-Klein modes.

\subsection{Moduli Spectroscopy}

At large volume, the single most important factor determining the moduli masses is
the stabilised internal volume. The different scales are suppressed compared to the 4-dimensional
Planck scale by various powers of the internal volume. In table \ref{Table1} we show this
scaling explicitly for the various moduli. There are also
model-dependent $\mc{O}(1)$ factors, which we do not show explicitly.
\begin{table}
\caption{Moduli spectrum for $\mbb{P}^4_{[1,1,1,6,9]}$ in terms of
  $\mc{V}_s^0 = \langle \mc{V}_s \rangle$}
\label{Table1}
\centering
\vspace{3mm}
\begin{tabular}{|c|c|}
\hline
Scale & Mass \\
\hline
\textrm{4-dimensional Planck mass} & $\frac{4 \pi \mc{V}_s^0}{g_s}
m_s = M_P$ \\
\textrm{String scale } $m_s$ & $m_s = \frac{g_s}{\sqrt{4 \pi \mc{V}_s^0}} M_P$ \\
\textrm{Stringy modes} $m_S$ & $2 \pi m_s = \frac{g_s \sqrt{\pi}}{\sqrt{\mc{V}_s^0}} M_P$ \\
\textrm{Kaluza-Klein modes} $m_{KK}$ & $\frac{2 \pi}{{\cal
    V}_s^{\frac{1}{6}}} m_s =
\frac{g_s \sqrt{\pi}}{(\mc{V}_s^0)^{\frac{2}{3}}} M_P$ \\
\textrm{Gravitino} $m_{3/2}$ & $\frac{g_s W_0}{\sqrt{\mc{V}_s^0}} m_s =  
\frac{g_s^2 W_0}{\sqrt{4 \pi} \mc{V}_s^0} M_P$ \\
\textrm{Dilaton-axion } $m_{\tau}$
& $\frac{g_s N}{\sqrt{\mc{V}_s^0}} m_s = \frac{g_s^2 N}{\sqrt{4 \pi} \mc{V}_s^0} M_P$
\\
\textrm{Complex structure moduli } $m_{\phi}$ &
$\frac{g_s N}{\sqrt{\mc{V}_s^0}} m_s = \frac{g_s^2 N}{\sqrt{4 \pi} \mc{V}_s^0} M_P$
\\
\textrm{`Small' K\"ahler modulus } $m_{\tau_4}, m_{\tilde{\tau_4}}$ & $\frac{
  W_0}{\sqrt{\mc{V}_s^0}} m_s = 
\frac{g_s W_0}{\sqrt{4 \pi} \mc{V}_s^0} M_P$
\\
\textrm{Modulinos } $m_{\tilde{\tau}}, m_{\tilde{\phi}}$ & $\frac{g_s
  W_0}{\sqrt{\mc{V}_s^0}} m_s = 
\frac{g_s^2 W_0}{\sqrt{4 \pi} \mc{V}_s^0} M_P$
\\
\textrm{`Large' volume modulus} $m_{\tau_5}$ & $\frac{g_s
  W_0}{\mc{V}_s^0} m_s = \frac{g_s^2 W_0}{\sqrt{4 \pi} (\mc{V}_s^0)^{\frac{3}{2}}} M_P$ \\
\textrm{Volume axion} $m_{b_5}$ & $\exp( - (\mc{V}_s^0)^{\twothirds}) M_P \sim 0 $ \\
\hline
\end{tabular}
\end{table}

This spectrum has several characteristic features. First, the string scale is hierarchically
smaller than the Planck scale. The internal volume depends exponentially on the inverse string coupling
and thus very small changes in the stabilised dilaton lead to large effects in the compact space.
A standard criticism of the large extra dimensions scenario was that it seemed very difficult to
obtain naturally from string theory; the above shows that this criticism
does not hold here.

The majority of moduli masses are stabilised at a high scale $\mc{O}\left(\frac{M_P}{\mc{V}_s^0}\right)$,
comparable to $m_{\frac{3}{2}}$ but below the scale of stringy and Kaluza-Klein modes.
There are two light moduli; the radion and its associated axion.
The latter has an extremely small mass that is light in any units. As an axion, one would like to
use this as a solution to the strong CP problem. Unfortunately this
axion corresponds precisely to the D7 gauge theory with gauge
coupling determined by the inverse size of the large 4-cycle which
then would be extremely small, and so we do not expect the Standard Model to
live on such branes. The radion may have cosmological implications
which it would be interesting to explore further.

The principal factor entering the scales is the internal volume, and we present in table
\ref{Table2} possible spectra arising from various choices of the
internal volume. We would like to emphasise that the reason we can
talk about `choices' of the volume is that its stabilised value is
exponentially sensitive to $\mc{O}(1)$ parameters such as the string
coupling, and thus it may be dialled freely from the Planck to TeV scales.
The moduli spectra are shown for GUT, intermediate
and TeV string scales. 
If the fundamental scale is at the GUT scale, all moduli are heavy
with the exception of the
light $b_5$ axion. For the intermediate scale, the volume modulus $\tau_5$ is relatively light. Its mass does not
present a problem with fifth force experiments but may be problematic in a cosmological context
\cite{hepph9308292, hepph9308325}. Finally, for TeV scale gravity all
moduli are very light.
In particular, $\tau_5$ is now so light ($10^{-17}$ eV) 
to be in conflict with fifth force experiments. It would then be 
difficult to realise this scenario unless either for some reason observable matter
did not couple to $\tau_5$ or its mass received extra corrections.

\begin{table}
\caption{Moduli spectra for GUT, intermediate and TeV string scales}
\label{Table2}
\centering
\vspace{3mm}
\begin{tabular}{|c|c|c|c|c|}
\hline
Scale & Mass & GUT & Intermediate & TeV \\
\hline
$M_P$ & $M_P$ & $2.4 \ti 10^{18}$ GeV &  $2.4 \ti 10^{18}$ GeV &  $2.4 \ti 10^{18}$ GeV\\
$m_s$ & $\frac{g_s}{\sqrt{4 \pi \mc{V}_s^0}} M_P$ & $1.0 \ti 10^{15}$ GeV & $1.0 \ti 10^{12}$ GeV& $1.0 \ti 10^{3}$ GeV\\
$m_S$ & $2 \pi m_s = \frac{g_s \sqrt{\pi}}{\sqrt{\mc{V}_s^0}} M_P$ & $6 \ti 10^{15}$ GeV
& $6 \ti 10^{12}$ GeV& $6 \ti 10^{3}$ GeV\\
$m_{KK}$ & $\frac{2 \pi m_s}{(\mc{V}_s^0)^{\frac{1}{6}}} =
\frac{g_s \sqrt{\pi}}{(\mc{V}_s^0)^{\frac{2}{3}}} M_P$ & $1.5 \ti 10^{15}$ GeV & $1.5 \ti 10^{11}$ GeV
& $0.15$ GeV \\
$m_{3/2}$ & $\frac{g_s^2 W_0}{\sqrt{4 \pi} \mc{V}_s^0} M_P$ &
$1.5\ti 10^{12}$ GeV & $1.5\ti 10^6$GeV & $1.5\ti 10^{-12} $GeV\\
$m_\tau$ & $\frac{g_s N m_s}{\sqrt{\mc{V}_s^0}} = \frac{g_s^2 N}{\sqrt{4 \pi} \mc{V}_s^0} M_P$ &
$1.5\ti 10^{12}$ GeV & $1.5\ti 10^6$GeV & $1.5\ti 10^{-12} $GeV\\
$m_{cs}$ & $\frac{g_s N m_s}{\sqrt{\mc{V}_s^0}} = \frac{g_s^2 N}{\sqrt{4 \pi} \mc{V}_s^0} M_P$ &
$1.5\ti 10^{12}$ GeV & $1.5\ti 10^6$GeV & $1.5\ti 10^{-12} $GeV\\
$m_{\tau_4}, m_{b_4}$ & $\frac{a_4 g_s W_0}{\sqrt{4 \pi} \mc{V}_s^0} M_P$  &
$1.5\ti 10^{11}$ GeV & $1.5\ti 10^5$GeV & $1.5\ti 10^{-11} $GeV\\
$m_{\tau_5}$ & $\frac{g_s^2 W_0}{\sqrt{4 \pi} (\mc{V}_s^0)^{\frac{3}{2}}} M_P$
& $2.2 \ti 10^{10}$ GeV & 22 GeV &  $2.2\ti 10^{-26}$ GeV\\
$m_{b_5}$ & $\exp( - a_5 \tau_5) M_P \sim 0 $ & $\sim 10^{-300}$ GeV &
$\exp( -10^6)$ GeV&
$\exp(-10^{18})$ GeV\\
\hline
\end{tabular}
\end{table}

\subsection{On the validity of the effective field theory}
\label{EFT}

Using these results we can go back and check that the
four-dimensional effective field theory is self-consistent.
Our use of an $\mc{N} =1 $ supergravity framework 
should be valid so long as there is a separation of scales in which
the 4-dimensional physics decouples from the high-energy physics.
Let us enumerate the consistency conditions that a candidate minimum must satisfy
in order to trust the formalism.
First,
\be
\label{Constraint1}
\langle \mc{V}_s \rangle = \langle \mc{V}_E g_s^{\frac{3}{2}} \rangle
\gg 1.
\ee
To control the $\alpha'$ expansion, the string-frame compactification
volume must be much greater than unity. It is important to keep track
of frames here; 
this condition is often incorrectly stated with an Einstein frame
$(g_{\mu \nu, s} = e^{\phi/2} g_{\mu \nu, E})$ volume. However, in Einstein frame the $\alpha'$
corrections come with inverse powers of $g_s$ (as in (\ref{alpha'correction})) and
thus the string frame volume is the correct measure.

Secondly, we require
\bea
\label{Constraint2}
 \langle V \rangle & \ll & m_s^4, \nonumber \\
\langle V \rangle & \ll & m_{KK}^4.
\eea
The vacuum energy must be suppressed compared to the string and
compactification scales. Otherwise, the neglect of stringy and Kaluza-Klein modes
in the analysis is untrustworthy. We likewise
require the particle masses to decouple, namely that
\be
\label{Constraint3}
m_{\frac{3}{2}}, m_{\phi}, m_{\tilde{\phi}} \ll m_s, m_{KK},
\ee
where $\phi$ and $\tilde{\phi}$ are generic moduli and modulini.

There is one further potential constraint we wish to discuss. The
$\mc{N}=1$ SUGRA energy can be written
\be
V_{\mc{N}=1} = \underbrace{e^\K\left[G^{i \bar{j}}D_i W D_{\bar{j}}
    \bar{W}\right]}_{\textrm{F-term energy}}
- \underbrace{3e^\K \vert W \vert^2}_{\textrm{gravitational energy}},
\ee
and we may thus define a susy breaking energy $m_{susy}$, by $m_{susy}^4 =
e^\K\left[G^{i \bar{j}} D_i W D_{\bar{j}} \bar{W}\right]$. In no-scale models, $m_{susy}^4$ cancels against
$3 e^\K \vert W \vert^2 $ to give a vanishing cosmological constant. Is it necessary to
require $m_{susy} \ll m_{KK}, m_s$? As $m_{susy} \sim
W_0 m_s$, imposing this would lead to the constraints
\bea
\label{MSusyConstraints}
W_0 & \ll & 1 \nonumber \\
W_0 \langle \mc{V}_s \rangle^{\frac{1}{3}} & \ll & 1.
\eea
Normally, there is a twofold reason for requiring $m_{susy} \ll m_s$.
First, once susy is broken the vacuum energy is of $\mc{O}(m_{susy}^4)$
and secondly, the boson-fermion mass splittings are $\mc{O}(m_{susy})$. However, neither
reasons are valid here. In a no-scale model, irrespective of the value of $m_{susy}$, the vacuum
energy vanishes. Furthermore, the boson associated with supersymmetry breaking is
massless in the no-scale approximation (as the potential is flat), whereas the associated fermion
is the goldstino that is eaten by the gravitino. It is the gravitino that sets the scale of
Bose-Fermi splitting, but this has mass $m_{3/2} = e^{\frac{\K}{2}}
\vert W \vert \sim \frac{m_{susy}}{\sqrt{{\cal{V}}}}$
at large volumes.
Requiring $m_{3/2} \ll m_{KK},
m_s$ leads to the much weaker constraint
\be
\label{WeakConstraint}
W_0 \ll \langle \mc{V}_s \rangle^{\frac{1}{3}}.
\ee
Thus $m_{susy}$ as defined above is an imaginary
scale; it sets neither the scale of Bose-Fermi splitting
nor the vacuum energy. Thus we shall only consider (\ref{Constraint1}),
(\ref{Constraint2}) and (\ref{Constraint3}) as relevant constraints.
We should note that this distinction only arises because we are in a
large extra dimensions scenario. If $\langle \mc{V}_s \rangle \sim
\mc{O}(1)$, the conditions (\ref{MSusyConstraints}) and
(\ref{WeakConstraint}) coalesce.

There is finally a consistency conditions peculiar to IIB
flux compactifications. If $\tau_i$ are the divisor volumes, we
require
\be
\label{Constraint4}
\frac{a_i \tau_{i,s}}{g_s} = a_i \tau_{i,E} \gg 1.
\ee
This allows us to neglect
multi-instanton contributions.

Let us now consider these constraints as applied to our model.
As the stabilised volume is exponentially large, (\ref{Constraint1}) is trivially
satisfied. The conditions (\ref{Constraint2}) depend on the vacuum energy
at the minimum. After the de Sitter uplift, these are satisfied by construction.
Before the uplift, we recall that the vacuum energy at the minimum is
$\mc{O}(\frac{W_0^2 M_P^4 g_s^4}{4 \pi {\cal{V}}^3})$,
whereas $m_{KK}^4 \sim \frac{\pi^2 g_s^4
  M_P^4}{{\cal{V}}^{\frac{8}{3}}}$. This gives a restriction
\be
W_0 \ll \pi^{\frac{3}{2}} \langle \mc{V}_s \rangle ^{\frac{1}{6}}.
\ee
At large volumes this is not an onerous condition to satisfy.

The particle mass constraints are likewise satisfied. The most dangerous of these is the requirement
$m_{3/2}, m_{\phi} \ll m_{KK}$. From (\ref{csKK}) we require
\be
\frac{g_s N}{2 \pi} \ll \langle \mc{V} \rangle^{\third},
\ee
were $N$ is a measure of the typical number of flux
quanta\footnote{Not to be 
confused with $N$ measuring the rank of the
  hidden sector group that enters in the coefficients $a_i$ for
  gaugino condensation potentials. } and can be
taken
to be $\mc{O}(\sqrt{\frac{\chi}{24}}) \sim 30$. At the large volumes we work
at this constraint is satisfied comfortably.
The constraints on the divisor volumes are also satisfied at large
volume, as for the 
`small' divisor $\tau_4$, $a_4 \tau_4 \sim \ln {\cal V}$ at the minimum.

Thus all consistency conditions are satisfied and we see no reason to
regard the use of $\mc{N} = 1$ supergravity as inconsistent.
An important point is that we obtain no strong constraints on the value of $W_0$;
this is in contrast to KKLT-type solutions, for which
very small values of $W_0$ are essential. As large values of $W_0$ are
preferred by the statistical results of Douglas and collaborators 
\cite{hepth0411183, hepth0404116, hepth0307049}, we expect a
typical solution to have large $W_0$. The maximum value of $W_0$ is
determined by the fluxes satisfying the tadpole conditions and can be
in general of order $10-100$.

In KKLT constructions the constraint (\ref{Constraint1}) leads to the
requirement $W_0 \ll 1$ and $g_s$ not too small. To be more precise,
as $A_i e^{- \frac{a_i \tau_i}{g_s}} \sim W_0$, this gives
$$
-\frac{g_s \ln (W_0)}{a_i} \gg 1.
$$
If this is satisfied then satisfying (\ref{Constraint2}) is automatic.
As in such compactifications $\langle \mc{V}_s \rangle^{\third} \sim
\mc{O}(1)$, (\ref{Constraint3}) gives 
$$
\frac{g_s N}{2 \pi} \ll \mc{O}(1).
$$
This is in general hard to satisfy (as pointed out in \cite{hepth0309170}).

\subsection{The general case}
\label{GenSection}
We have so far focused on a particular model,
${\mathbb{P}}^4_{[1,1,1,6,9]}$. Let us now argue that the above framework will in
fact be valid for all Calabi-Yaus with $h^{2,1} > h^{1,1}$, assuming
the existence of appropriate non-perturbative superpotentials.

In the ${\mathbb{P}}^4_{[1,1,1,6,9]}$ example, we found that of
the two K\"ahler moduli $\tau_4, \tau_5$ one ($\tau_5$) was large,
whereas the other ($\tau_4$) was stabilised
at a small value. Our claim is that this behaviour will persist
in the general case,
with  only one K\"ahler modulus large and all others tending to be
small. Therefore there is only one modulus responsible for the
large volumes obtained.

To argue this, let us write the large-volume expression for the scalar potential as: 
\be 
\label{LargeToSmallEquation}
V = \ \frac{C_1 e^{-{a_i\tau_i}-
a_j\tau_j}}{\cal V}\ - \ \frac{C_2
  e^{-a_i\tau_i}}{{\cal V}^2}\ +\ \frac{C_3}{{\cal V}^3},
\ee
with the factors $C_i$ depending on powers of the K\"ahler moduli at
the most. $C_3 \sim - \chi(M)$ and so it is  positive as long as
$h^{2,1} > h^{1,1}$.
We shall not quibble here over frames or the factors of $g_s$ in the
exponent
as they do not affect the argument.

Let us start in a position where $\mc{V} \gg 1$, $V < 0$ and there are
many large moduli $\tau_i \gg 1$.
We may investigate the behaviour of the potential as one of the 
$\tau_i$ fields change. Originally, all terms non-perturbative in the
large $\tau_i$ can be neglected. As the second term is the only negative term,
it wants to increase its magnitude in order to
minimise the value of the potential.
This will naturally reduce the value of the corresponding $\tau_i$
in order for the exponential to be more relevant. As
long as the other large moduli are adjusted to keep $\mc{V}$ constant,
this reduces the value of the potential. This continues until  $e^{-a_i \tau_i}
\sim \frac{1}{\mc{V}}$, when the (positive) double exponential term in
(\ref{LargeToSmallEquation}) becomes important and the modulus
will be stabilised.
 
We may carry on doing this with all the large moduli, but since we are holding
$\cal V$ constant, one (combination)
of the fields must remain large. This finally leaves us with a picture of a
manifold with one large 4-cycle (and corresponding 2-cycle), but all
other cycles of size close to the string scale.

In order to get a clearer picture
let us make a brief geometric digression. The volume of a Calabi-Yau
can be expressed either in terms of 2-cycles, $\mc{V} = \frac{1}{6} k_{ijk}t^i t^j
t^k$, or 4-cycles $\mc{V} = \mc{V}(\tau_i)$. Then the matrix
\be
\label{d2Vrelation}
M_{ij} = \frac{\partial^2 \mc{V}}{\partial \tau_i \partial \tau_j}
\ee
has signature $(1,h^{1,1} - 1)$ (one plus, the rest minus). This
follows from the result that $\tau_i = \tau_i (t^j)$ is simply a
coordinate
change on K\"ahler moduli space, and it is a standard result
\cite{CandelasDeLaOssa} that
\be
M_{ij}^{'} = \frac{\partial^2 \mc{V}}{\partial t^i \partial t^j}
\ee
has signature $(1, h^{1,1} -1)$. The coordinate change cannot change
the signature of the metric.

This signature manifests itself in explicit models. In both
the ${\mathbb{P}}^4_{[1,1,1,6,9]}$ example studied above and an $\mc{F}_{11}$
model also studied in \cite{hepth0404257}, the volume may in fact be
written explicitly in terms of the divisor volumes. With the $\tau_i$ as
defined in \cite{hepth0404257}\footnote{Note that here the $\tau_i \neq 
\frac{\partial \mc{V}}{\partial t_i}$, but are rather linear
combinations thereof.}, we have
\bea
\label{ModelVolumes}
{\mathbb{P}}^4_{[1,1,1,6,9]} & & {\cal{V}} = {1\over{9\sqrt{2}}}\left(\tau_5^{\frac{3}{2}} -
\tau_4^{\frac{3}{2}}\right), \nonumber \\
& & \tau_4 = \frac{t_1^2}{2} \textrm{ and } \tau_5 = \frac{(t_1 + 6t_5)^2}{2}.
\nonumber \\
\mc{F}_{11} & & {\cal{V}} =
\frac{1}{3 \sqrt{2}} \left( 
2\left(\tau_1+\tau_2
+2\tau_3\right)^{3/2}-\left(\tau_2+2\tau_3\right)^{3/2} - \tau_2^{3/2}
\right), 
\nonumber \\
& & \tau_1 = \frac{t_2}{2}\left(2 t_1 + t_2 + 4 t_3\right), \,
\tau_2 = \frac{t_1^2}{2}, \textrm{ and } \tau_3 = t_3 \left(t_1 + t_3\right).
\eea
For the $\mc{F}_{11}$ model, from the expressions for
$\tau_i$ in terms of 2-cycles, we may see that it is consistent to have
$\tau_1$ large and $\tau_2, \tau_3$ small but not otherwise. The signature of
$d^2 V$ is manifest in (\ref{ModelVolumes}); each expression contains
$h^{1,1} - 1$ minus signs. There is another important point. In each case, there is
a well-defined limit in which the overall volume goes to infinity and all but one divisors remain
small. These limits are given by $(\tau_5 \to \infty, \tau_4 \textrm{ constant })$ and
$(\tau_1 \to \infty, \tau_2, \tau_3 \textrm{ constant })$ respectively. Furthermore,
in each case this limit is unique: e.g. the alternative limit $(\tau_2 \to \infty, \tau_1, \tau_3
\textrm{ constant })$ is not well-defined.

\begin{figure}[ht]
\begin{center}
\makebox[10cm]{
\epsfxsize=12cm
\epsfysize=8cm
\epsfbox{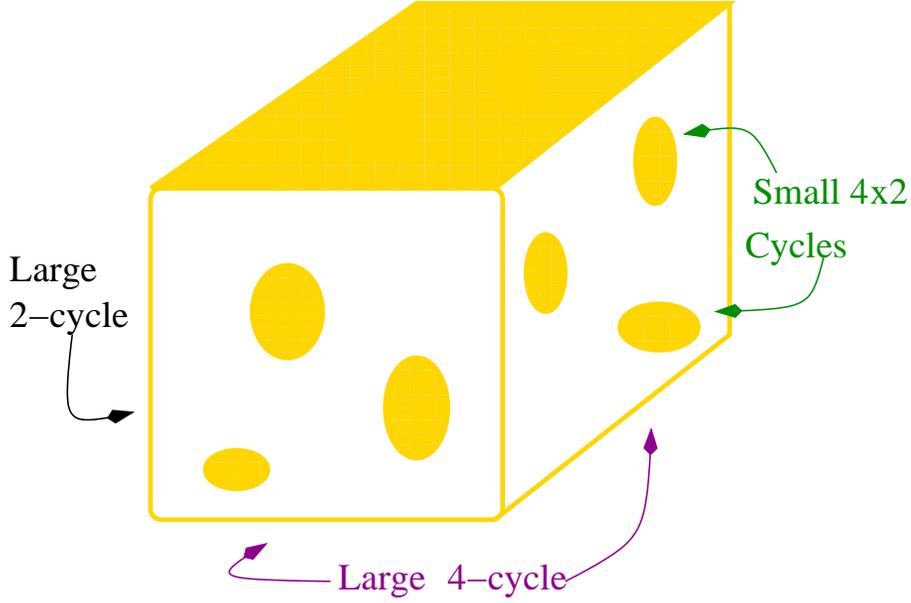}}
\end{center}
\caption{A Swiss cheese picture of a Calabi-Yau. There is one pair of
large 2- and 4-cycles - increasing the cycle volume increases the overall
volume. The other pairs are such that increasing the cycle volume decreases the
overall volume.}
\label{SwissCheese}
\end{figure}

This motivates a `Swiss-cheese' picture of the Calabi-Yau, illustrated in figure \ref{SwissCheese}.
A Swiss cheese is a 3-manifold with 2-cycles. Of these 2-cycles, one ($t_b$) is `large'
and the others ($t_{s,i}$) are small. The volume of the cheese can be written
\be
\mc{V} = t_b^{3/2} - \sum_i t_{s,i}^{3/2},
\ee
and $\frac{\partial^2 \mc{V}}{\partial t_i \partial t_j}$ has signature $(1, h^2 - 1)$.
The small cycles are internal; increasing their volume decreases the overall volume of the manifold.
There is one distinguished cycle that controls the overall volume; this cycle may be
made arbitrarily large while holding all other cycles small, and controls the overall volume.
For all other cycles, an arbitrary increase in their volume decreases the overall volume and eventually
leads to an inconsistency. The small cycles may be thought of as local effects; if the bulk
cycle is large, the overall volume is largely insensitive to the size of the small cycles.

To capture the above, let us consider a Calabi-Yau with divisors $\tau_b, \tau_{s,i}$ such that the
volume can be written
\be
\label{CYLvol}
{\cal{V}} = \left(\tau_b + \sum a_i \tau_{s,i}\right)^{\frac{3}{2}} -
 \left(\sum b_i \tau_{s,i}\right)^{\frac{3}{2}}
- .... - \left(\sum k_i \tau_{s,i}\right)^{\frac{3}{2}}.
\ee
We assume that a limit $\tau_b \gg\tau_{s,i}$ is well-defined.
By working in this limit, the minus signs can be seen to follow from (\ref{d2Vrelation}). The form given above is
valid globally for both ${\mathbb{P}}^4_{[1,1,1,6,9]}$ and
$\mc{F}_{11}$ models. 
The form (\ref{CYLvol}) is illustrative and it is not important for
our argument that it hold generally; the important assumption is that
there exists a well-defined limit $\tau_b \gg \tau_{s,i}$.
We also note that the argument that follows can then be recast as in \cite{hepth0502058}
using expressions solely in terms of 2-cycle moduli.
In the limit $\tau_b \to \infty, \tau_{s,i} \textrm{ small }$,
the scalar potential takes the form
\be
V = e^\K \left[ G^{i \bar{\jmath}} \partial_i W \partial_{\bar{\jmath}} \bar{W}
+ G^{i \bar{\jmath}} ((\partial_i \K) W \partial_{\bar{\jmath}} \bar{W} + c.c.)
+ \frac{3 \xi}{4 \mc{V}} \right]
\ee
We take $W$ to be
\be
W = W_0 + \sum_{s,i} A_i e^{- a_i T_i}.
\ee
We may include an exponential dependence on $\tau_b$ in $W$; as
$\tau_b \gg 1$ this is in any case insignificant. Now,
\be
G^{i \bar{\jmath}} \partial_i \K \propto \tau_j,
\ee
and we may also verify that, so long as $i$ and $j$ both correspond to small moduli,
\be
G^{i \bar{\jmath}} \sim \mc{V} \sqrt{\tau_s^{'}}
\ee
where $\tau_s^{'}$ is a complicated function of the $\tau_{s,i}$ that scales
linearly under $\tau_{s,i} \to \lambda \tau_{s,i}$. The scalar potential then takes the form
\be
V = \frac{\sqrt{\tau_s^{'}} \partial_i W \cdot \partial_j W}{\mc{V}} -
\frac{\tau_i \cdot \partial_i W}{\mc{V}^2} + \frac{3 \xi}{4 \mc{V}^3}.
\ee
If we then take the decompactification limit $\tau_b \to \infty$ with
$\partial_i W = \frac{1}{\mc{V}}$ and $\tau_i \sim \mc{O}(\ln \mc{V})$,
then the potential goes to zero from below.
As this result is independent of the strength of the non-perturbative corrections,
which always eventually dominate the positive $\alpha'^3$ terms, there is an associated
minimum at large volume.

The geometric interpretation of this is that the `external' cycle controlling the
overall volume may be very large, whereas the small, `internal' cycles are
always stabilised at small volumes. As in section \ref{MassesSection}, the
masses associated with the
moduli parametrising the small cycles appear at a high scale
with $m_{\tau_i} \sim \mc{O}(\frac{g_s^2 W_0}{\mc{V}})$, comparable to the masses
of the dilaton and complex structure moduli. Thus the resulting spectrum is largely
model-independent - the moduli associated with small internal cycles take masses at
a high scale and decouple from the low energy physics. However, the
`volume' modulus, which is distinguished and model-independent, is relatively light and may be
of cosmological importance.

Even though we think we have explored a generic case there is still
room for different pictures to emerge. In particular, we may imagine
a Calabi-Yau for which ${\cal V}= t_1 t_2^2 + F(t_n)$ with $t_n \neq
t_1,t_2$. Then $\tau_1 =t_2^2$ and $\tau_2=2t_1 t_2$, and so if we
had $\tau_2$ large and $\tau_1$ small, we would have $t_1$ large
and $t_2$ small. This will give a throat-like picture where one
throat has a large two-cycle and small four-cycle whereas the other
throat has a large four-cycle and a small two-cycle, with all other
cycles small. This would give the
interesting possibility of making contact with the $0.1$ mm fundamental scale
scenario since then ${\cal V} \sim t_1$ and, in the throat, there are only two large
dimensions \cite{add, susha}. This may be of phenomenological interest
given the potential to search for deviations of gravity at the
submillimeter scale and the connection between this scale and dark
energy \cite{susha}. However, we do not know if there are Calabi-Yau manifolds
with this property.

Another possibility would be the existence of Calabi-Yaus for which the
limit $\tau_b >> \tau_{s,i}$ is not well-defined. In that case, it
would not be possible to realise large volumes without having several
large divisors rather than just one. This does not hold for
$\mbb{P}^4_{[1,1,1,6,9]}$ or $\mc{F}_{11}$; if it were to hold for
other models it would be interesting to study the consequences for the
above mechanism.

\section{Soft Supersymmetry Breaking Terms}
We have so far described the spectrum of the closed  moduli sector of the
theory.
In a typical string model we will have, besides these fields, the open
string moduli, usually corresponding to the location of D-branes, and
matter fields living on D-branes. For
phenomenological purposes, it is this spectrum that has more relevance.
In this section we will study the soft supersymmetry breaking
Lagrangian corresponding to the large volume minimum we have found.

It is worth noticing that this sector is more model dependent and
depends on how the standard model is embedded within this string
theory construction: for a given Calabi-Yau flux compactification we
may have several different ways to embed the standard model.

Let us concentrate on the two main possibilities by considering matter
fields coming from D3 or D7 branes.

\subsection{The moduli-matter couplings}

In \cite{hepth031232} the general K\" ahler potential for Calabi-Yau
orientifolds with D3 branes was derived by a dimensional reduction. The result is
\begin{equation}
\label{kahlerpot}
K(S, T, U,\phi) = -\log (S+\bar{S}) -\log(-i\int
\Omega\wedge \bar{\Omega}) - 2 \log ({\cal{V}} (T, U, \phi)).
\end{equation}
Here $U$ are the complex structure moduli
and $\phi^i, i=1,2,3$ are the scalar fields corresponding to the
position of a stack of D3 branes on the Calabi-Yau. The Calabi-Yau volume $\cal{V}$
is to be understood
as a function of the complexified K\" ahler moduli $T$, the expression
for which is:
\begin{equation}
\label{complexified}
T_\alpha = \tau_\alpha + i\rho_\alpha + i\mu_3 l^2
(\omega_\alpha)_{\imath{\bar{\jmath}}} {\rm Tr} \phi^i
\left( \bar{\phi}^{\bar{\jmath}}
- {i\over2} \bar{U}^{\hat{a}} (\bar{\chi_{\hat{a}}})^{\bar{\jmath}}_l \phi^l\right).
\end{equation}
$\omega_\alpha$ are a basis for (1,1) forms on the Calabi-Yau which
survive the orientifold projection, while $\chi_{\hat{a}}$ form a basis
for $(2,1)$ forms that have negative sign under the orientifold
projection. Their number equals the number of complex structure
moduli $U^{\hat{a}}$ surviving the orientifold projection. $\mu_3=(2\pi)^{-3} \alpha'^{-2}$ 
is the RR charge of a single D3 brane.  Also $l=2\pi\alpha'$ and so
$\mu_3 l^2 = 1/(2\pi).$
For orientifold compactifications with both D3 and D7 branes, the K\" ahler potential was derived
in \cite{hepth0409098} and has the form
\begin{equation}
\label{KahlerWithD7}
\K(S,T, U, \zeta) = -\log (S+\bar{S}+2i \mu_7 {\cal{L}}_{AB} \zeta^A
\bar{\zeta}^{\bar{B}}) -\log (-i\int\Omega\wedge \bar{\Omega})-2 \log ({\cal{V}} (T, U,\phi)),
\end{equation}
where ${\cal{L}}_{AB}$ are certain geometric quantities and $\zeta^A$ are
the moduli describing the position of the D7 brane. We are neglecting for now the possibility
of Wilson line moduli.
\subsection{F-terms}
We will use the standard formalism for calculating soft supersymmetry breaking terms,
as described for example in \cite{hepph9707209}. We proceed by expanding
the K\" ahler potential and superpotential in terms of the visible sector fields $\varphi$
\begin{eqnarray}
\K &=& \hat{\K} + (\tilde{\K}_{i\bar{\jmath}}) \varphi^i
\bar{\varphi}^{\bar{\jmath}} + Z_{ij} \varphi^i \varphi^j + \cdots,
\nonumber\\
W &=& \hat{W} + \mu_{ij} \varphi^i \varphi^j + Y_{ijk} \varphi^i \varphi^j \varphi^k +\cdots,
\end{eqnarray}
where $\hat{\K}, \tilde{\K}_{i\bar{\jmath}}, Z_{ij}, \mu_{ij}$ and $Y_{ijk}$
depend on the hidden moduli only.
F-term supersymmetry breaking in a hidden sector is
characterised by nonvanishing expectation values for the auxiliary fields of the hidden
sector chiral superfields. These F-terms may be written as
\begin{equation}
\label{fterms}
{\bar{F}}^{\bar{m}} = e^{{\hat{\K}/{2M_P^2}}} {\hat{\K}}^{\bar{m}n} {{D_n \hat{W}}\over{M_P^2}}.
\end{equation}
In this and all subsequent formulae, $m$ and $n$ range over the
hidden moduli - the dilaton, complex structure and K\" ahler moduli
in our case.
We henceforth work in Planck mass units and do not include explicit
factors of $M_P.$

Given the F-terms, the various soft parameters can be calculated.
For example, assuming a diagonal matter field metric
$\tilde{\K}_{i\bar{\jmath}}=\delta_{i\bar{\jmath}}
\tilde{\K}_i$\footnote{$\tilde{\K}_i$ is not to be confused with the derivative of
the K\" ahler potential with respect to modulus $i.$}, the masses squared of
canonically normalised matter fields $\varphi^i$  can be written as
\begin{equation}
\label{softmass}
m_i^2 = m_{3/2}^2 +V_0 - F^{m} {\bar{F}}^{\bar{n}} \partial_m \partial_{\bar{n}} \log \tilde{\K}_i,
\end{equation}
where $V_0$ denotes the value of the cosmological constant.
The normalised gaugino masses for a sector with gauge kinetic
function $f_a$ are
\begin{equation}
\label{gauginomass}
M_a = {1\over2} ({\rm Re} f_a)^{-1} F^m \partial_m f_a,
\end{equation}
while the A-terms of normalised matter fields $\hat{\varphi}^i$ are $A_{ijk}
\hat{Y}_{ijk} \hat{\varphi}^i \hat{\varphi}^j \hat{\varphi}^k$ with
\begin{eqnarray}
\label{aterms}
A_{ijk} &=& F^m (\hat{\K}_m +\partial_m \log Y_{ijk}- \partial_m \log (\tilde{\K}_i
\tilde{\K}_i\tilde{\K}_k)),\nonumber\\
\hat{Y}_{ijk} &=& Y_{ijk} {{\bar{\hat{W}}}\over{|\hat{W}|}} e^{\hat{\K}/2}
(\tilde{\K}_i \tilde{\K}_j \tilde{\K}_k)^{-1/2}.
\end{eqnarray}
Finally, if $Z_{ij} = \delta_{ij} Z$ and $\mu_{ij} = \mu \delta_{ij}$,
the B-term $\hat{\mu} B \hat{\varphi}^i \hat{\varphi}^i$ for the field
$\varphi^i$ can be written as
\begin{eqnarray}
\hat{\mu} B &=& (\tilde{\K}_i)^{-1} \Bigl\{ {\bar{\hat{W}}\over{|\hat{W}|}} e^{\hat{\K}/2}
\mu \Bigl[F^m (\hat{\K}_m +\partial_m \log\mu - 2\partial_m \log
(\tilde{\K}_i))\nonumber\\
& &{}-m_{3/2}\Bigr]+(2m_{3/2}^2+V_0) Z - m_{3/2} \bar{F}^{\bar{m}}
\partial_{\bar{m}} Z \nonumber\\
& &{}+ m_{3/2} F^m (\partial_m Z - 2 Z\partial_m \log
(\tilde{\K_i}))\nonumber\\
& &{}- \bar{F}^{\bar{m}} F^n (\partial_{\bar{m}} \partial_n
Z - 2\partial_{\bar{m}} Z \partial_n \log (\tilde{\K}_i))\Bigr\}.
\end{eqnarray}
where the effective $\mu$-term is given by
\begin{equation}
\hat{\mu} = \left( {\bar{\hat{W}}\over{|\hat{W}|}} e^{\hat{\K}/2} \mu + m_{3/2} Z
 - F^{\bar{m}} \partial_{\bar{m}} Z \right) (\tilde{\K}_i)^{-1}.
\end{equation}
We shall set $V_0=0$ since soft masses are to be evaluated after lifting
the vacuum energy.

\subsubsection{Relative size of F-terms}
It is important to have an idea of the approximate sizes of the various
F-terms used in the computation of soft terms. We shall again work in the
context of the ${\mathbb{P}}^4_{[1,1,1,6,9]}$ model with two K\" ahler moduli $T_4,T_5.$
At large volume, the relevant parts of the inverse metric are
\begin{eqnarray}
\label{invmetric}
\K^{\bar{T}_4 T_4} &\sim& {\cal{V}},\nonumber\\
\K^{\bar{T}_4 T_5} &\sim& {\cal{V}}^{2/3},\nonumber\\
\K^{\bar{T}_5 T_5} &\sim& {\cal{V}}^{4/3},\nonumber\\
\K^{\bar{S}   T_4} &\sim& {1\over{{\cal{V}}}},\nonumber\\
\K^{\bar{S}   T_5} &\sim& {1\over{{\cal{V}}^{1/3}}}.
\end{eqnarray}
The derivatives of the K\" ahler potential are
$\partial_4 \K \equiv \partial_{T_4} \K \sim 1/{\cal{V}}$ and $\partial_5 \K\sim
{\tau_5^{1/2}/{\cal{V}}} \sim 1/{\cal{V}}^{2/3}.$
Since at the minimum of the scalar potential $D_i W=0$ for
the dilaton and complex structure moduli, the volume dependence
of F-terms is given by
\begin{eqnarray}
F^4 &\sim& {1\over\cal{V}}\left(
{\cal{V}\over\cal{V}} +
{{\cal{V}}^{2/3}\over {\cal{V}}^{2/3}}\right) \sim {1\over{\cal{V}}} \nonumber\\
F^5 &\sim& {1\over{\cal{V}}}\left( {{\cal{V}}^{4/3}\over {\cal{V}}^{2/3}} +
{{\cal{V}}^{2/3}\over {\cal{V}}} \right)\sim {1\over{\cal{V}}^{1/3}}\nonumber\\
F^S &\sim& {1\over{\cal{V}}} \left( {1\over{{\cal{V}}}} +
{1\over{{\cal{V}}^2}}\right) \sim {1\over{\cal{V}}^2}.
\end{eqnarray}
We see that, at large volume, ${\cal{V}}^{-1/3} \sim F^5\gg F^4 \sim {\cal{V}}^{-1}\gg F^S
\sim {\cal{V}}^{-2}.$

The F-terms corresponding to complex structure moduli vanish since
$\K^{\bar{U}i}=0$ for $i$ ranging over the dilaton and K\" ahler moduli,
even after including $\alpha'$ corrections, and also $D_U W=0$ at
the minimum of the scalar potential if we only turn on ISD
fluxes.

\subsection{D3 branes}
\subsubsection{Scalar masses}
\label{D3MassSection}
To calculate the masses of D3 moduli it is sufficient to work with
a low energy theory only containing K\" ahler moduli and D3 moduli,
since ISD fluxes do not give masses to D3 moduli whereas they
give large masses (${\cal{O}} (m_{3/2})$) to the dilaton, the complex
structure moduli and the D7 brane moduli (\cite{hepth0105097,hepth0311241,hepth031232}).
We will also restrict to a single D3-brane rather than a stack, and 
assume that the D3 moduli metric is diagonal for simplicity
of calculations --- this is of course probably untrue for the
Calabi-Yau under consideration, but since we are primarily concerned
with the volume scaling of the soft parameters, the features we obtain
ought to be quite generic.

Equations (\ref{kahlerpot}) and (\ref{complexified}) determine the way D3 moduli 
enter the K\" ahler potential.
Concentrating on a single D3 modulus $\phi$, the K\" ahler potential after integrating
out the dilaton and complex structure moduli becomes
\begin{equation}
\K = -2\log \Bigl[
(T_5 + \bar{T}_5 - c|\phi|^2)^{3/2} - (T_4+\bar{T}_4 - d|\phi|^2)^{3/2}
+ {\xi'\over2}\Bigr] + \K_{cs} +2\log(36),
\end{equation}
Here $\xi'=36\xi$ and  $c, d$ parametrise our ignorance of the forms
$\omega_\alpha$ and are expected to be ${\cal{O}} (1).$ 
This is obtained from the original
K\" ahler potential $\K = -2\log \left({\cal{V}} +{\xi\over2}\right) + \K_{cs} =
-2\log \left(\tau_5^{3/2} - \tau_4^{3/2} +{9\sqrt{2}\xi\over2}\right) + 2\log \left(9\sqrt{2}\right).$

The superpotential is
\begin{equation}
\label{wnodilaton}
W = \hat{W} = {g_s^2\over{\sqrt{4\pi}}} \left(W_0 + A_4 e^{- {a_4\over g_s} T_4} + A_5 e^{- {a_5\over
    g_s} T_5}\right),
\end{equation}
there being no supersymmetric $\mu$-term for D3 brane scalars. We
will assume for simplicity of expressions that $W$ is real.

After expanding $\K$ around $\phi=0$ we get the following
expressions for $\tilde{\K}_i$ and $\hat{\K}$:
\begin{eqnarray}
\label{ktilde}
\tilde{\K}_i &=& 3{{{c(T_5+\bar{T}_5)^{1/2} - d(T_4+\bar{T}_4)^{1/2}}}\over
{(T_5+\bar{T}_5)^{3/2} - (T_4+\bar{T}_4)^{3/2}+{\xi'\over2}}},\nonumber\\
\hat{\K} &=& -2\log \bigl[ (T_5+\bar{T}_5)^{3/2} - (T_4+\bar{T}_4)^{3/2}+{\xi'\over2}\bigr].
\end{eqnarray}
We now introduce the variables $X=(T_4+\bar{T}_4)^{1/2},
Y=(T_5+\bar{T}_5)^{1/2}$ to simplify various expressions appearing
in the rest of this section.

It is important to note that in the no-scale approximation (obtained
by setting $\xi=0$ in $\K$ and $A_4=A_5=0$ in $W$), the nonvanishing F-terms
are $F^4 = -e^{\hat{\K}/2} X^2 W$,
$F^5 = -e^{\hat{\K}/2} Y^2 W$ (this is derived in
appendix C). We also have, after a redefinition of $c$ and $d$,
\begin{equation}
\log (\tilde{\K}_i) = \log (cX+dY)-\log ({\cal{V}}) + {const}.
\end{equation}
It is easy to check that $F^m \bar{F}^{\bar{n}}
\partial_m \partial_{\bar{n}} \log (cX+dY) = -e^{\hat{\K}}|W|^2/2.$ Also
\begin{equation}
F^m \bar{F}^{\bar{n}}\partial_m
  \partial_{\bar{n}} \log ({\cal{V}}) =
-(1/2) \hat{\K}_{m\bar{n}} F^m \bar{F}^{\bar{n}} =
-(3/2)e^{\hat{\K}}|W|^2
\end{equation}
in this approximation, so that
\begin{equation}
F^m \bar{F}^{\bar{n}} \partial_m \partial_{\bar{n}} \log {\tilde{\K}_i}
= e^{\hat{\K}}|W|^2.
\end{equation}
This cancels against $m_{3/2}^2 = e^{\hat{\K}} |W|^2$ in the
expression for soft masses (\ref{softmass}) giving the no-scale result
$m_i^2=0.$

Let us estimate the size of soft scalar masses in the ${\mathbb{P}}^4_{[1,1,1,6,9]}$ 
model, without doing an explicit calculation.
Consider the expression (\ref{softmass}). The inclusion of $\alpha'$
and nonperturbative effects alters the F-terms (through $\hat{\K}$, $\hat{\K}^{\bar{T}_i T_j}$
and $\partial_i W$) and the expression for $\tilde{\K}_i.$
After including nonperturbative contributions (but temporarily neglecting $\alpha'$
corrections) the F-terms are
\begin{eqnarray}
F^4 &\sim& {1\over{\cal{V}}} + {1\over{\cal{V}}} \K^{T_4 \bar{T}_4}
(\partial_4 W)
\nonumber\\
F^5 &\sim& {1\over{{\cal{V}}^{1/3}}} + {1\over{\cal{V}}} \K^{T_4 \bar{T}_5}
(\partial_4 W).
\end{eqnarray}
By construction, the modulus $\tau_4$ is small at the minimum,
while $\tau_5$ is exponentially large, so we are justified in including
only the nonperturbative contribution from $\tau_4.$ Moreover, $\tau_4$
is such that at the (AdS) minimum we have
\begin{equation}
-\partial_{\tau_4} W = {{a_4 A_4}\over{g_s}} e^{-{a_4\over{g_s}} \tau_4}\sim {\xi^{1\over3}
|W_0|\over{\langle{\cal{V}}_s\rangle}},
\end{equation}
and so $(\partial_4 W)\sim 1/{\cal{V}}.$ Therefore, we may use the
expressions for the inverse metric $\K^{\bar{T}_i T_j}$ given in
(\ref{invmetric}) to write
\begin{eqnarray}
F^4 &\sim& {1\over{\cal{V}}} + {1\over{\cal{V}}},
\nonumber\\
F^5 &\sim& {1\over{{\cal{V}}^{1/3}}} + {1\over{{\cal{V}}^{4/3}}}.
\end{eqnarray}
In these expressions the second term corresponds to the modification coming
from the nonperturbative addition to the superpotential.
The dominant contributions to scalar masses will come from
terms of the form
$F^m_{np} \bar{F}^{\bar{n}} \partial_m \partial_{\bar{n}} \log (\tilde{\K}_i)$
where $F^m_{np}$ is the nonperturbative contribution to the F-term.
Finally we have to see how $\partial_m \partial_{\bar{n}} \log
\tilde{\K}_i$ scales with the volume, for $m,n \in \{ 4,5\}.$
Using the explicit expressions (\ref{logcu}) derived in the
appendix we see that
\begin{eqnarray}
\label{logcunew}
\partial_4 \partial_4 \log (cX+dY) &\sim& {\cal{V}}^{-1/3},
\nonumber\\
\partial_4 \partial_5 \log (cX+dY) &\sim& {1\over{\cal{V}}},
\nonumber\\
\partial_5 \partial_5 \log (cX+dY) &\sim& {\cal{V}}^{-4/3}.
\end{eqnarray}
Using the above expressions for F-terms and (\ref{logcunew}), we have
\begin{eqnarray}
F^4_{np} \bar{F}^{\bar{4}} \partial_4 \partial_{\bar{4}} \log (cX+dY)
&\sim& {1\over{\cal{V}}} {1\over{\cal{V}}}  {\cal{V}}^{-1/3}
= {\cal{V}}^{-7/3},\nonumber\\
F^4_{np} \bar{F}^{\bar{5}} \partial_4 \partial_{\bar{5}} \log (cX+dY)
&\sim& {1\over{\cal{V}}} {1\over{{\cal{V}}^{4/3}}}  {\cal{V}}^{-1}
= {\cal{V}}^{-10/3},\nonumber\\
F^5_{np} \bar{F}^{\bar{5}} \partial_5 \partial_{\bar{5}} \log (cX+dY)
&\sim& {1\over{{\cal{V}}^{4/3}}} {1\over{{\cal{V}}^{1/3}}}  {\cal{V}}^{-4/3}
= {\cal{V}}^{-3}.
\end{eqnarray}
We therefore expect the masses squared to be ${\cal{O}}( 1/{\cal{V}}^{7/3}).$
A similar analysis can be done for the terms of type
$F^m_{np} \bar{F}^{\bar{n}} {\hat{\K}}_{m\bar{n}}$ and also including $\alpha'$
corrections - these turn out to be subleading compared to
the contribution from nonperturbative corrections to the superpotential.
The explicit expression for moduli masses can be found in
appendix C.

Our conclusion is that the masses of D3 moduli are suppressed with respect
to the gravitino mass, by a factor ${\cal{V}}^{-1/6}$ (the factors
of $g_s$ and $W_0$ also present are derived in appendix C):
\begin{equation}
m_i = {\cal{O}} (1) {g_s^2 W_0\over{\sqrt{4\pi} ({\cal V}^0_s)^{7/6}}}.
\end{equation}

\subsubsection{Gaugino masses}
We next turn our attention to gaugino masses for D3 branes.
The gauge kinetic function is $f=\mu_3 l^2 S= S/(2\pi)$ and the normalised
gaugino masses
\begin{equation}
M_{D3} = {1\over2} 2\pi ({\rm Re} S)^{-1} F^S = {{\cal{O}}} \left(
{1\over{{\cal{V}}^2}}\right).
\end{equation}
$F^S$ must be calculated using the full K\" ahler
potential, before integrating out the dilaton and complex structure moduli.
$M_{D3}$ turns out to be proportional to $g_s^2$ and
$W_0$ in the same way scalar masses are---this can be deduced from the
$1/g_s$ prefactor in the inverse metric $\K^{\bar{S} T_i}$
(as shown in \cite{hepth0412239}) and the factor of $g_s^{3/2}$ in
$W$, which can be observed in (\ref{Potentials}).

\subsubsection{A-terms}
We can also estimate the magnitude of the A-terms which
vanish in the no-scale approximation.
If we use (\ref{ktilde}) in the A-term expression (\ref{aterms}), with
constant $Y_{ijk}$, we obtain
\begin{eqnarray}
A_{\phi\phi\phi} &=& e^{\hat{\K}/2}\Bigl\{ -{{3\xi}\over{4\cal{V}}}W + (\partial_4 W)
\Bigl[
{X^2\over{36{\cal{V}}}} (2Y^3 + X^3 - {\xi'\over2}) - {3Y\over{\cal{V}}} X^2 Y^2
 \nonumber\\
& & +{3\over{2(cX+dY)}}\left( {c\over3} (2Y^3 + X^3 -{\xi'/2}) +
d X^2 Y\right)\Bigr]\Bigr\}.
\end{eqnarray}
As $\partial_4 W\sim {\cal{O}} (1/{\cal{V}})$,
$A\sim Y^2/{{\cal{V}}^2} \sim {\cal{V}}^{-4/3}.$
Similarly to the scalar and gaugino masses, the dependence of $A$ on $g_s$ and $W_0$ is given by
by $A\propto g_s^2 W_0.$

\subsubsection{$\mu$-terms and B-terms}
For D3 branes, the supersymmetric $\mu$ term vanishes, but there is an effective $\mu$-term
generated by the Giudice-Masiero mechanism \cite{GiudiceMasiero} due to the appearance of
a bilinear in the K\" ahler potential dependent on complex
structure moduli, as follows from formulae (\ref{kahlerpot}) and
(\ref{complexified}).
The prefactor of the bilinear $\phi^i \phi^j$ in the expansion of the
K\" ahler potential is
\begin{equation}
Z_{ij} = {{3\mu_3 l^2}\over{ {\cal{V}} +\xi/2}} t^{\alpha}
(\omega_\alpha)_{(j|\bar{s}} (\bar{\chi}_{\hat{a}})^{\bar{s}}_{|i)}
\bar{U}^{\hat{a}}.
\end{equation}
For simplicity we consider only one $Z$ with $Z=(c'X+d'Y)(a_i U^i)/(
  {\cal{V}} + \xi/2).$
The complex structure moduli dependence of $Z$ is in fact 
unimportant since the F-terms corresponding to these moduli are
vanishing.
The calculation of the effective $\mu$-term will be very similar to the
computation of A-terms; we do not need to differentiate by $U$, and
can absorb $a_i U^i$
into $c'$ and $d'$ to write $Z=(c'X+d'Y)/({\cal{V}} + \xi/2).$
Then 
\begin{eqnarray}
\partial_4 Z &\sim& (1/{\cal{V}}),\nonumber\\
\partial_5 Z &\sim&  1/{\cal{V}}^{4/3}
\end{eqnarray}
so that $F^m \partial_m Z$ gives rise to terms
\begin{eqnarray}
e^{\K/2} \K^{44} (\partial_4 W)\partial_4 Z &\sim& 1/{\cal{V}}^2,\nonumber\\
e^{\K/2} \K^{54} {\partial_4 W}\partial_5 Z &\sim& 1/{\cal{V}}^{8/3}.
\end{eqnarray}
Assuming the complex structure moduli to be fixed at ${\cal{O}}
(1)$ values, it is easy to confirm that the $\hat{\mu}$ term scales as
${\cal{O}} (1/{\cal{V}}^{4/3}).$

As $Z$ can be treated as
$(c'X+d'Y)/ ({\cal{V}}+\xi/2)$ and by analogy with the calculation
of the masses squared, it is easy to see that the expression
for $\hat{\mu} B$ behaves like ${\cal{O}} (1/{\cal{V}}^{7/3}).$

Not including anomaly mediated contributions, the dependence of D3
brane soft terms 
on ${\cal{V}}, W_0$ and $g_s$ is summarised in table \ref{d3softtable}.

\begin{table}
\label{d3softtable}
\caption{Soft terms for D3 branes
  (AMSB contributions not included)}
\centering
\vspace{3mm}
\begin{tabular}{|c|c|c|c|c|}
\hline
Scale & Mass & GUT & Intermediate & TeV \\
\hline
\textrm{Scalars} $m_i$ & $\frac{g_s^2}{({\cal{V}}_s^0)^{7/6}}W_0 M_P$ &
$3.6\ti 10^{11}$ GeV & $3.6\ti 10^4$GeV & $3.6\ti 10^{-17}$GeV\\
\textrm{Gauginos} $M_{D3}$ & $\frac{g_s^2}{({\cal{V}}_s^0)^2}W_0 M_P$ &
$3.6\times 10^9$GeV & $3.6\ti 10^{-3}$GeV & $3.6\ti 10^{-39}$ GeV\\
\textrm{A-term} $A$ & $\frac{g_s^2}{({\cal{V}}_s^0)^{4/3}}W_0 M_P$ &
$3.2\times 10^{11}$GeV
& $3.2 \ti 10^3$ GeV  & $3.2\ti 10^{-21}$GeV\\
$\mu$-\textrm{term} $\hat{\mu}$ & $\frac{g_s^2}{({\cal{V}}_s^0)^{4/3}}W_0
M_P$ & $3.2\times 10^{11}$GeV
& $3.2 \ti 10^3$ GeV  & $3.2\ti 10^{-21}$GeV\\
\textrm{B term} $\hat{\mu}B$ &
 $\frac{g_s^2}{({\cal{V}}_s^0)^{7/6}}W_0 M_P$ &
$3.6\ti 10^{11}$ GeV & $3.6\ti 10^4$GeV & $3.6\ti 10^{-17}$GeV\\
\hline
\end{tabular}
\end{table}

\subsection{D7 branes}

The open string sector of D7 branes can give rise to several different
types of moduli in the low energy theory.
There are geometric moduli corresponding to deformations of the
internal 4-cycle $\Sigma$ that the D7 brane wraps.
As discussed in \cite{hepth0409098}, the number of such moduli is
related to the $(2,0)$ cohomology of
the cycle $\Sigma.$ There are also Wilson line moduli $a_I$
which will be present if the cycle $\Sigma$ possesses harmonic $(1,0)$
forms. However this is not the case for most Calabi-Yaus. If
present, these enter the K\" ahler potential through the complexified K\"
ahler moduli, which are further redefined from (\ref{complexified}) to
\begin{equation}
\label{complexified2}
T_\alpha = \tau_\alpha + i\rho_\alpha + i\mu_3 l^2
(\omega_\alpha)_{\imath{\bar{\jmath}}} {\rm Tr} \phi^i
\left( \bar{\phi}^{\bar{\jmath}}
- {i\over2} \bar{U}^{\hat{a}} (\bar{\chi_{\hat{a}}})^{\bar{\jmath}}_l
\phi^l\right)
+ \mu_7 l^2 C^{I\bar{J}} a_I \bar{a}_{\bar{J}},
\end{equation}
for geometry dependent coefficients $C^{I\bar{J}}.$
The D7 geometric moduli are generically given large, ${\cal{O}} (m_{3/2})$
masses after turning on fluxes---
this is most easily seen from the F-theory perspective, where the
D7 moduli are among the complex structure moduli of the Calabi-Yau 4-fold $M_8$
of the F-theory compactification. The relevant Gukov-Vafa-Witten
superpotential is
\begin{equation}
W = \int_{M_8} G_4\wedge\Omega,
\end{equation}
which generically induces a nontrivial potential for the D7 moduli.
We expect the $\alpha'$ and nonperturbative effects not
to affect the already large D7 scalar masses by very much.

For the case of a single geometric D7 brane modulus, the K\" ahler
potential is 
$-\log (S+\bar{S} - {\cal{L}} |\zeta|^2)$ giving $\tilde{\K} = {\cal{L}}/(S+\bar{S}).$
As $F^S$ vanishes before breaking
the no-scale structure, for D7 branes
$F^m \bar{F}^{\bar{n}} \partial_m \partial_{\bar{n}} \log {\tilde{\K}}$ vanishes and
the flux-induced mass of $\zeta$ is ${\cal{O}} (m_{3/2})$ (which is of order $1/{\cal{V}}$).
Including $\alpha'$ corrections, $F^S$ is no longer zero and
\begin{equation}
F^S \bar{F}^{\bar{S}} \partial_S \partial_{\bar{S}} (-\log (S+\bar{S}))
= {1\over{(S+\bar{S})^2}} F^S \bar{F}^{\bar{S}} =
{\cal{O}} \left( {1\over{{{\cal{V}}^4}}}\right).
\end{equation}
This is a manifestly tiny correction to $m_{\zeta}$, and so
\begin{equation}
m_{\zeta} \approx m_{3/2} = {\cal{O}} (1) {g_s^2 W_0 \over{\sqrt{4\pi}{\cal V}^0_s}}M_P.
\end{equation}

The masses of Wilson line moduli for D7 branes (if present)
may be found by using the modified K\" ahler coordinates (\ref{complexified2}).
Since these moduli appear in the K\" ahler potential in a similar
fashion to D3 moduli, 
the calculation of their masses squared will be exactly parallel. 
We therefore expect the Wilson line moduli to obtain ${\cal{O}}
({\cal{V}}^{-7/6})$ 
masses due to nonperturbative effects:
\begin{equation}
m_{Wilson} = {\cal{O}} (1) {g_s^2 W_0\over{\sqrt{4\pi} ({\cal V}^0_s)^{7/6}}} M_P.
\end{equation}

For D7 branes, the gauge kinetic function $f_a$ is given by the K\"
ahler modulus, $T_a$, of the cycle the D-branes wrap:
\begin{eqnarray*}
f_a = {T_a\over{2\pi}}.
\end{eqnarray*}
In general $F^{T^a}\ne0$ and the gaugino masses are nonvanishing.
For the $\mathbb{P}^4_{[1,1,1,6,9]}$ example, if $a=4$ then
\begin{equation}
M_4 = \pi F^4/({\rm Re } \ T_4) \sim {\cal{V}}^{-1},
\end{equation}
and if $a=5$ we get 
\begin{equation}
M_5 \sim {\cal{V}}^{-1/3}/ {\cal{V}}^{2/3} = {\cal{V}}^{-1}.
\end{equation}
In either case
\begin{equation}
M_{D7} = {\cal{O}} (1) {g_s^2 W_0 \over{\sqrt{4\pi} {\cal V}^0_s}}.
\end{equation}

Another crucial difference between D3 and D7 branes is that a supersymmetric
$\mu$-term can be induced for geometric D7 moduli with only ISD fluxes;
in fact, it was shown in \cite{hepth0408036} that for vanishing magnetic fluxes on the
D7 brane, the $\mu$-term corresponds to the $(2,1)$ component of flux
$G_3.$ 
This will give an extra contribution to the mass of D7 geometric
moduli, to be added to the SUSY breaking contribution proportional to the $(0,3)$
flux component. However, for simplicity, we only assume the latter to
be present.

Let us now consider the A-term corresponding to a D7 scalar field $\phi$
with $\tilde{\K}_i=\tilde{\K}_j=\tilde{\K}_k={\cal{L}}/(S+\bar{S}).$
We have
\begin{equation}
A_{\phi\phi\phi} = F^4 (\partial_4 \hat{\K}) + F^5 (\partial_5 \hat{\K}) + F^S
(\partial_S \hat{\K}) + {3F^S\over{S+\bar{S}}}.
\end{equation}
As $\partial_S \hat{\K} = -1/(S+\bar{S}) + {\cal{O}} ({\cal{V}}^{-1})$ and
$F^S\sim {1/{\cal{V}}^2}$, the largest contribution to
the D7 A-term comes from $F^5 (\partial_5 \K)$,and so
\begin{equation}
A\sim F^5 (\partial_5 K) \sim {1\over{{\cal{V}}^{1/3}}}\cdot {1\over{{\cal{V}}^{2/3}}} = {1\over{{\cal{V}}}}.
\end{equation}
Thus the A-term is ${\cal{O}} (m_{3/2}).$

If we try and realise the Standard Model on D7 branes, there is one
potential worry. The Yang-Mills gauge coupling is determined by
$g_{YM,a}^{-2} = \textrm{Re } f_a$, and thus if $f_a = T^5$ then $g_{YM}$ 
would be unacceptably small. However, in general, as shown in section
\ref{GenSection},
we have $T_b \gg 1$ but $T_{s,i}$ relatively small. Thus so long as
the Standard Model is realised on branes wrapping the smaller cycles,
the resulting gauge kinetic function will have phenomenologically
acceptable values near $1$.

\begin{table}
\label{d7softtable}
\caption{Soft terms for D7 branes (AMSB not included)}
\centering
\vspace{3mm}
\begin{tabular}{|c|c|c|c|c|}
\hline
Scale & Mass & GUT & Intermediate & TeV \\
\hline
\textrm{Scalars} $m_\zeta$ & $m_{3/2}$ &
$1.5\ti 10^{12}$ GeV & $1.5\ti 10^6$GeV & $1.5\ti 10^{-12} $GeV\\
\textrm{Gauginos} $M_4,M_5$ & $m_{3/2}$ &
$1.5\ti 10^{12}$ GeV & $1.5\ti 10^6$GeV & $1.5\ti 10^{-12} $GeV\\
\textrm{A-term} $A$ & $m_{3/2}$ & 
$1.5\ti 10^{12}$ GeV & $1.5\ti 10^6$GeV & $1.5\ti 10^{-12} $GeV\\
$\mu$-\textrm{term} $\hat{\mu}$ & $m_{3/2}$ & 
$1.5\ti 10^{12}$ GeV & $1.5\ti 10^6$GeV & $1.5\ti 10^{-12} $GeV\\
\textrm{B term} $\hat{\mu}B$ &
$m_{3/2}$ &  $1.5\ti 10^{12}$ GeV & $1.5\ti 10^6$GeV & $1.5\ti 10^{-12} $GeV\\
\hline
\end{tabular}
\end{table}

\subsection{D3-D7 strings}
The intersections of D3 and D7 branes can give rise to massless open
string states. Unfortunately their appearance in the K\" ahler
potential of the low energy theory cannot be deduced from the dimensional
reduction of Dirac-Born-Infeld and Chern-Simons actions for stacks for
branes, and thus their soft terms are difficult to analyse from
the four-dimensional point of view. However, in \cite{hepth0408036}, 
the soft masses for 3-7 scalar fields due to a general
flux background were computed using various symmetry arguments. 
It was found that no scalar or fermion masses are generated.
In some particular
examples where it is known how the 3-7 scalars enter the 4D K\" ahler
potential this can be seen from the four dimensional perspective. In
\cite{hepph9812397, hepth0406092} the K\" ahler potential was obtained
for compactifications on $T^2\ti T^2\ti T^2$ with D3 and D7 branes. 
The dependence on 3-7 fields $\phi_{37}$ is
\begin{equation}
\K = { {|\phi_{37}|^2}\over{(T_1+\bar{T_1})^{1/2} (T_2+\bar{T_2})^{1/2}}} +\cdots,
\end{equation}
for a D7 brane wrapping the first two tori. If there is only one overall K\" ahler
modulus $T=T_1=T_2$ this becomes
\begin{equation}
\K = {|\phi_{37}|^2 \over{T+\bar{T}}}+\cdots,
\end{equation}
and the same no-scale cancellation argument of section \ref{D3MassSection}
for D3 matter applies for $\phi_{37}.$

As for D3 matter, after including no-scale breaking effects, we expect
the masses of 3-7 fields to be
\begin{equation}
m_{37} = {\cal{O}} (1) {g_s^2 W_0\over{\sqrt{4\pi} ({\cal V}^0_s)^{7/6}}}.
\end{equation}

Although not done explicitly, the A-terms for D3-D7 matter could be
computed by  setting some of the $\tilde{K}$ to equal ${\cal{L}}/(S+\bar{S})$ in formula (\ref{aterms}).
\subsection{D-terms and de Sitter lifting}

\subsubsection{D-terms and Soft Supersymmetry Breaking}

To perform a semirealistic computation of the soft supersymmetry
breaking terms, we need to uplift the nonsupersymmetric AdS vacuum
obtained by fixing the moduli using $\alpha'$ corrections and
nonperturbative effects. This can be done in several ways, all of which
share some essential features. One option, put forward in
\cite{hepth0301240} is to use an anti-D3 brane
at the bottom of a highly warped Klebanov-Strassler throat
to generate the uplifting term. Although the resulting uplifting term is
reminiscent of a D-term in the low energy theory, supersymmetry is
broken explicitly, albeit by a small amount.  An alternative means of uplifting to
de Sitter was proposed in \cite{hepth0309187}, where one turns on
magnetic fluxes on a D7 brane wrapping a compact 4-cycle in the Calabi-Yau.
The uplifting term thus generated can indeed be interpreted as a
D-term in the low energy theory \cite{hepth0502059}. Again, one needs the brane to be in a
highly warped region to be able to fine tune the resulting
cosmological constant.
Yet another mechanism for producing an uplifting term was proposed
in \cite{hepth0402135}: instead of having a strongly warped region, one can
look for local minima of the no-scale potential and find one with
$V_0\ne 0$ (for a contrary view, see \cite{dealwis}). This will happen for certain non-ISD choices of fluxes;
for example, in a model with only one K\" ahler modulus $T$, setting
$W=\int G_3\wedge \Omega$ gives a source term for $T$
\begin{equation}
V_0 = {1\over{(S+\bar{S})(T+\bar{T})^3}}\left| \int G_3^* \wedge \Omega \right|^2.
\end{equation}
$V_0$ will be nonzero if we turn on a non-ISD $(3,0)$
component in $G_3.$ The large number of complex structure moduli and
hence choices of fluxes generically present in Calabi-Yau
compactifications should enable us to find a $G_3$ with sufficiently
small $\int G_3^* \wedge \Omega.$ Finally, one could also use the
ideas of \cite{hepth0402088}.

Irrespective of the uplift mechanism used,
extra contributions to scalar masses will appear.
For concreteness let us use an uplift appropriate to IASD fluxes
\begin{equation}
D={\epsilon\over{{\cal{V}}^2}} =
{\epsilon\over\bigl[
(T_5+\bar{T}_5-c|\phi|^2)^{3/2} - (T_4+\bar{T}_4-d|\phi|^2)^{3/2}\bigr]^2},
\end{equation}
expressed in terms of the complexified K\" ahler moduli $T_4$ and $T_5.$
This may be expanded around $\phi=0$ as
\begin{equation}
D = {\epsilon\over\bigl[(T_5+\bar{T}_5)^{3/2}-(T_4+\bar{T}_4)^{3/2}\bigr]^2}
\left( 1+{3|\phi|^2(c(T_5+\bar{T}_5)^{3/2}-d(T_4+\bar{T}_4)^{3/2})
\over{(T_5+\bar{T}_5)^{3/2}-(T_4+\bar{T}_4)^{3/2}}}\right).
\end{equation}
Ignoring $\alpha'$ corrections, the coefficient of $|\phi|^2$ in the brackets can be
identified with the prefactor of kinetic term for the $\phi$ fields,
(\ref{ktilde}), and so
the uplift gives a contribution $\epsilon/{\cal{V}}^2$ to scalar
masses squared. 
To uplift to Minkowski space we require $\langle D \rangle = \langle
V_{min} \rangle\sim {\cal{O}} (1/{{\cal{V}}^3})$ which gives
$\epsilon\sim 1/{\cal{V}}.$
Thus the uplift contribution to scalar masses squared is ${\cal{O}}
(1/{\cal{V}}^3)$,
much less than the ${\cal{O}} ({\cal{V}}^{-7/3})$ obtained from
nonperturbative and $\alpha'$ effects.

If IASD fluxes are present\footnote{We thank S. Kachru for
  reminding us of this possibility.}, we no longer have $F^U = 0$.
However, if the IASD fluxes are to serve as a lifting term, we can
  estimate the magnitude of the complex structure F-terms. We require
$$
\frac{e^{\K_{cs}} \K_{cs}^{\bar{\imath}\jmath} (D_{\bar{\imath}} W)
  (D_j W)}{\mc{V}^2}
\sim \frac{1}{\mc{V}^3},
$$
and so $D_i W\sim {\cal{V}}^{-1/2}$ It then follows that $F^U \sim
\mc{V}^{-\frac{3}{2}}$
and so the resulting effect on the masses is subleading to the F-terms
associated with the K\"ahler moduli.

\subsubsection{Uplifting to de Sitter Space and Numerical Estimates}

We wish to investigate whether the AdS minima of \cite{hepth0502058}
can be lifted to de Sitter ones using a D-term and how that changes
the values of the K\" ahler moduli at the minimum. The lifting
potential is assumed to be of the form
\begin{equation}
V_{{\rm uplift}} = {\epsilon\over{{\cal{V}}^{4/3}}}.
\end{equation}
This corresponds to the result $\epsilon/T^2$ in \cite{kklmmt},
where $\epsilon$ is essentially equal to the exponentially small
factor $e^{-8\pi K/(3 g_s M)}$
at the bottom of the Klebanov-Strassler type throat and $T$ is an
overall radial modulus.
The value of the scalar potential at the AdS minimum is ${\cal{O}}
(1/{{\cal{V}}^3})$ so that we want $\epsilon$ to be of order
${\cal{O}} (1/{{\cal{V}}^{5/3}}).$
Concentrating on our 2-modulus toy model, we expect that the
values of $\tau_4$ and $\cal{V}$ at the minimum of the scalar potential
should not change by too much after lifting to de Sitter. To see this,
recall that the scalar potential before including the D-term
roughly has the form
\begin{equation}
\label{scalarpot}
V = {\lambda \sqrt{\tau_4} e^{-2a_4 \tau_4}\over{\cal{V}}}
- {\mu\over{{\cal{V}}^2}} \tau_4 e^{-a_4 \tau_4} + {\nu\over{{\cal{V}}^3}}.
\end{equation}
At the minimum one has
\begin{equation}
{\cal{V}} = {\cal{V}}_0 = {\mu\over\lambda} \sqrt{\tau_4} e^{a_4 \tau_4}
\left( 1\pm \sqrt{1-{{3\nu\lambda}\over{\mu^2 \tau_4^{3/2}}}} \right).
\end{equation}
Also, in the approximation $a_4\tau_4\gg1$, we have
\begin{equation}
\tau_4 = \left( {{4\nu\lambda}\over{\mu^2}}\right)^{2/3}.
\end{equation}
Thus if we add an uplift term of the form ${\epsilon/{{\cal{V}}^{4/3}}}$
with $\epsilon \sim 1/{{\cal{V}}_0^{5/3}}$,
near ${\cal{V}}={\cal{V}}_0$ we can write the total scalar potential as in (\ref{scalarpot}) but with
$\nu$ shifted to $\nu+\epsilon{{\cal{V}}_0}^{5/3}.$
To see how this works in practice, we pick the values
$a_4 = 2\pi/10, A_4 = 1, g_s = 1/10, W_0=10, \xi=1.31$, giving $\tau_4 = 4.17$
and ${\cal{V}}_0 = 4.5\ti 10^{10}.$ We include a D-term
with $\epsilon = 4.9\ti 10^{-19}$ and get a de Sitter minimum
with $\tau_4 = 4.213$ and ${\cal{V}} = 5.9\ti 10^{10}$ and with
value of the scalar potential $2.5\ti 10^{-35}.$
Note that $\epsilon \cdot {{\cal{V}}_0}^{5/3} \approx 0.28$ so
the numerics agree with our estimates from the analytic formulae.
Also note that since $\cal{V}$ depends on $\tau_4$ exponentially,
an ${\cal{O}} (1)$ change in $\tau_4$ can induce a change in $\cal{V}$
of almost an order of magnitude.

From the size of $\epsilon$ necessary to lift the cosmological
constant to zero, we can also esimate the size of the
warping at the bottom of the Klebanov-Strassler throat: recall the D-term is
$e^{4A_{min}}/{{\cal{V}}^2}$ which has to cancel against
${\cal{O}} (1/{{\cal{V}}^3})$, so that  $e^{A_{min}}\sim {\cal{V}}^{-1/4}.$
In our numerical example this is roughly $2\ti 10^{-3}$ -
not particularly large.

The gravitino mass is $1.1\ti 10^5$GeV, while the scalar masses
for D3 moduli will be of the order ${\cal{V}}^{-7/6}$, which is
$1.8$TeV.
This number may be further lowered due to warping if the standard
model branes reside in the strongly warped region -
all masses at the bottom of the warped throat are supressed by $e^{A_{min}}.$

One more important question we can analyse numerically is how difficult it is to obtain
each of the scales (GUT, intermediate, TeV) in the two moduli model. The size of the volume
is largely determined by the ratio $a_4/g_s$, as can be seen from (\ref{VolumeAtMin}).
For gaugino condensation on D7 branes,
\begin{equation}
a_4 = {2\pi\over{N}}
\end{equation}
so that it is really the product $g_s N$ which determines $\cal{V}$ at the minimum. Fixing
 $g_s$ to be $1/10$, we can calculate the $N$ required to obtain each of the three string
scales. The results are shown in table \ref{ScalesAndN}.

\begin{table}
\label{ScalesAndN}
\caption{Ranks of gaugino condensation gauge groups required to obtain various string scales}
\centering
\vspace{3mm}
\begin{tabular}{|c|c|c|c|}
\hline
Scale & ${\cal V}_s$ &  $g_s N$ & $N$ if $g_s=0.1$\\
\hline
GUT & $4600$ & 2.25 & $22$\\
Intermediate & $4.6\ti 10^9$ & 0.85 & $9$\\
TeV & $4.6\ti 10^{27}$ & 0.30 & $3$\\
\hline
\end{tabular}
\end{table}

\subsection{Comparison with Anomaly mediated contributions}
In a general model with hidden sector SUSY breaking,
scalar masses, gaugino masses and A-terms are generated through loop effects
as a consequence of the super-Weyl anomaly (\cite{hepth9810155}). Assuming that soft terms
are generated solely through anomaly mediation, their values are
\begin{eqnarray}
M &=& {\beta_g\over{g}} m_{3/2},\nonumber\\
m_i^2 &=& -{1\over4}\left( {\partial \gamma_i \over \partial g}\beta_g
+ {\partial \gamma_i \over \partial y}\beta_y\right) m_{3/2}^2,\nonumber\\
A_y &=& -{\beta_y\over y}m_{3/2},
\end{eqnarray}
for gaugino mass $M$ and scalar masses $m_i.$
Here $\beta$ are the relevant $\beta$ functions, $\gamma_i$ are
the anomalous dimensions of the chiral superfields, while $g$ and $y$
denote the gauge and Yukawa couplings, respectively.
Alternatively one can use the compensator field formalism and write
\begin{eqnarray}
M &=& {{b g^2}\over{8\pi^2}} {F^C\over C_0},\nonumber\\
A_{ijk} &=& -{\gamma_i+\gamma_j+\gamma_k\over{16\pi^2}} {F^C\over C_0},
\nonumber\\
m_i^2 &=& {1\over{32\pi^2}} {d\gamma_i\over{d\log\mu}}\left|{F^C\over C_0}
\right|^2
+ {1\over{16\pi^2}} \left( \gamma^i_m F^m \left( \bar{{F^C\over C_0}}\right)\right),
\end{eqnarray}
where $C$ is the compensator field and $\gamma^i_m \equiv
\partial_m\gamma^i.$
We expect $F^C/C_0\sim m_{3/2}^2$
The contribution to scalar masses from AMSB is then (see \cite{hepth0208123})
\begin{equation}
m \sim b_0 \left( {g^2\over{16\pi^2}}\right) m_{3/2},
\end{equation}
where $b_0$ is the one loop beta function coefficient --- this
is suppressed with respect to the gravitino mass by a factor
$1/(16\pi^2).$ 
Similarly the gaugino masses are
suppresed by $1/(8\pi^2).$
As the nonperturbatively generated scalar masses ${\cal{O}} (m_{3/2}/{\cal{V}}^{7/6})$,
the AMSB contribution starts to
compete with the nonperturbative superpotential contribution to the
masses squared at the string scale of about $m_s = 2\ti 10^{10}$
GeV. The possibility of competition of the two contributions in the
context of the KKLT scenario was analysed in {\cite{choi}}.

This seems interesting, since scenarios incorporating the MSSM
with purely AMSB generated masses suffer from a common problem ---
negative masses squared for the scalar superpartners. In our case,
the ${\cal{V}}^{-7/6}$ contribution seems to have the ability to cure
the problem, but it might lead to difficulties with FCNC, if the
contribution is generation dependent. The nonperturbative contributions
to gaugino masses are generically very small and can only start
competing with AMSB induced masses at string scales of order
$5\ti 10^{14}$GeV or above.

\section{Differences  with KKLT Vacua}

Before concluding let us compare our results with
the more standard KKLT vacua.

\begin{enumerate}

\item First, as pointed out in section 3, the range of validity of
the KKLT effective action is very limited because the perturbative
corrections to the K\"ahler potential tend to dominate both at
large and small volume. This is due to the original no-scale
property of the K\"ahler potential. The KKLT potential is only valid
in the regime where the flux-induced tree level superpotential
is comparable to or less than its non-perturbative corrections.
Otherwise, the perturbative corrections dominate and $\alpha'$
corrections must be included to trust any results. Including these
corrections, the resulting 
potential includes both the large-volume minimum discussed above and,
for very small values of $W_0$, the KKLT minimum. When $W_0 \ll 1$ the
two minima coexist, and decreasing $W_0$ causes the two minima to
approach each other and to eventually merge. This behaviour is illustrated in figures
\ref{kkltbbcq1} and \ref{kkltbbcq2}.

\begin{figure}[ht]
\linespread{0.2}
\begin{center}
\epsfxsize=0.75\hsize \epsfbox{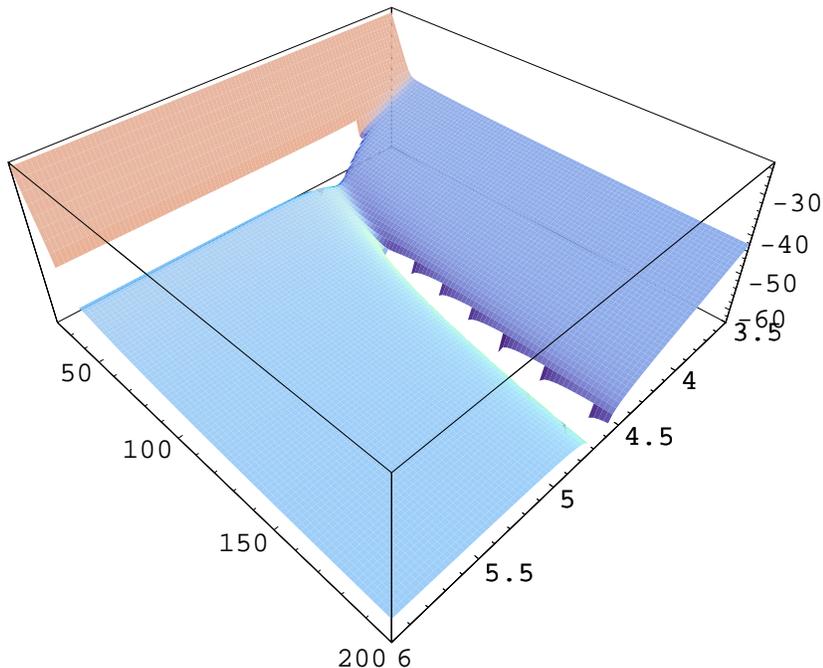}
\end{center}
\caption{ A plot of $\ln{V}$, showing the region of the scalar potential in which the large volume
  minimum coexists with a KKLT minimum at smaller volume. This picture
  is valid for very small $W_0$ ($\sim 10^{-10}$)}
\label{kkltbbcq1}
\end{figure}

\begin{figure}[ht]
\linespread{0.2}
\begin{center}
\epsfxsize=0.75\hsize \epsfbox{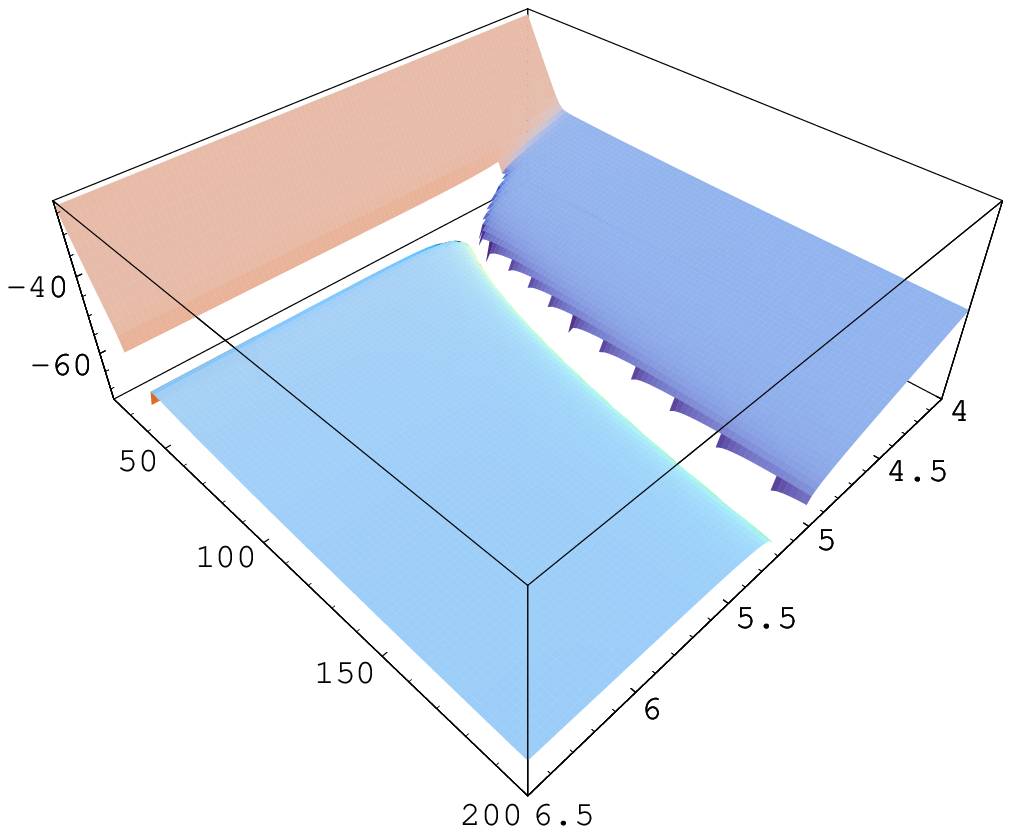}
\end{center}
\caption{ A plot of $\ln{V}$, showing how as $W_0$ is further
  decreased the large volume
  minimum merges with the KKLT minimum.}
\label{kkltbbcq2}
\end{figure}

\item In the KKLT approach it is common for the potential
  to develop tachyonic
directions after fixing the K\"ahler moduli. The reason
was explained in \cite{hepth0502058}. The dilaton ($S$) and complex
  structure ($U$) moduli are fixed by a term $\frac{DW \cdot DW (S,
  U)}{\mc{V}^2}$,
but the minimum of the potential is at $-\frac{e^{\K_{cs}} \vert W
  \vert ^2 (S,U)}{\mc{V}^2}$. As the volume scaling of these terms is the same, 
the negative part of the potential may 
trigger a destabilization of one or more directions.

For the large-volume minimum, the minimum of the potential 
is at $-\frac{e^{\K_{cs}} \vert W \vert^2 (S,U)}{{\cal V}^3}$. As the
contribution from dilaton and complex structure F-terms is still
positive and of $\mc{O}(\frac{1}{\mc{V}^2})$, movement of these moduli
from $D_i W = 0$.
could only increase 
the potential.
Therefore unlike KKLT there are
no tachyonic directions in the geometric moduli. This
is the main reason the models above are far simpler to analyse regarding
soft supersymmetry breaking. In the KKLT scenario, it
has so far not been possible to minimise the full potential, without following the two step procedure
in which dilaton and complex structure moduli are
fixed by the fluxes and then integrated out. This procedure
fails in many cases giving rise to tachyonic directions.
This was explicitly seen in \cite{hepth0411066, ciqs} for the simplest
cases with no complex structure moduli.\footnote{In this case however,
  a simple modification of the system can be made that stabilises the
  moduli \cite{ciqs}.
Non
  perturbative effects from D3 branes can induce a superpotential of the
  form $e^{-bS}$ which can be added to the KKLT superpotential, so 
$W=W_0 + Ae^{-aT} + B e^{-bS} $. This  gives
  rise to a minimum on the full $S,T$ plane. Adding D3 branes in
  principle adds new moduli corresponding to the position of the D3
  branes but these are also fixed as it can be seen that the
  corresponding masses for D3 brane scalars are positive.}. For more complicated
cases, the explicit dependence of the K\"ahler potential on the
complex structure moduli must be written in expansions around
the Gepner or conifold point and many terms are needed in order to
check that the full matrix of second derivatives has only positive
eigenvalues.

These tachyons have been analysed using statistical techniques in
\cite{hepth0404116, hepth0411183} where estimates have been given for
the number of tachyon-free vacua. We note that the tachyons do not
destabilise the supersymmetric AdS solution, which is protected by
supersymmetry, but do prevent an uplifting to a non-susy solution.

 \item As emphasised in \cite{hepth0502058}, given fixed $g_s$, the gravitino mass is
essentially independent of the value of $W_0$. This is relevant
for the recent interest in the scale of supersymmetry breaking in
landscape scenarios. This differs from the KKLT scenario,
in which the gravitino mass depends linearly on $W_0$ and is
a statistical variable, even at fixed $g_s$.

\item The main difference lies in the supersymmetry breaking
effects. In KKLT, the $AdS$ minimum is supersymmetric.
and the entire source of supersymmetry breaking terms is in the
uplift mechanism. The lifting term
dominates the soft supersymmetry breaking terms in the sense that
all $F$-terms vanish if this term is absent. In our scenario the
original minimum is already non-supersymmetric $AdS$, and
the sources of supersymmetry
breaking are principally the K\"ahler moduli $F$-terms. 
As discussed in section 5, these give the dominant contribution 
to soft terms irrespective of the uplift mechanism.

\end{enumerate}

\section{Conclusions and Outlook}

We have initiated a detailed phenomenological analysis of a large class of
Calabi-Yau models with all geometric moduli fixed. The moduli spectrum
and D3/D7 soft terms, both computed here, are the starting
point for any phenomenology of realistic models.

One remarkable result is that the volume is fixed at an exponentially
large value, which is the first time that large extra dimensions have been
realised in string theory. Furthermore, we have argued that in the general
case, there is one
dominant large K\"ahler modulus whereas the others are smaller with
sizes near the string scale (although large enough to neglect,
e.g. higher instanton corrections).
By large we mean large enough to obtain naturally, for instance,
 the electroweak scale from string theory (and thus solving the
hierarchy problem dynamically). Indeed, it is possible to obtain even 
larger extra dimensions corresponding to string scales below the electroweak scale. 
A possible way of getting submillimetre scales as in the six-dimensional
large extra dimensions scenarios was mentioned in section 4 but this
remains to be obtained in an explicit compactification. 
 In any case 
electroweak or intermediate string scales are in some sense easier to obtain
than larger scales, since smaller values of $N$ are required. 
However, all scales from the TeV to the Planck mass
are possible, although as mentioned above a TeV string scale suffers
from a volume modulus sufficiently light to be in conflict with fifth
force experiments. The main tuning
parameter
($g_s N$ for the two-moduli example) only need vary within a range of
$0.3-3$ to cover all physically accessible string
scales. For a value of $g_s \sim 0.1$ we need $N=9$
to have an intermediate string scale and $N=3$ to have the TeV scale
(where $N$ is the rank of the gauge group). 
Thus in these models there is no hierarchy problem. It is
interesting that for fixed $g_s$ arbitrarily small scales are not really accessible,
since we need $N\geq 2$ to have non-trivial gauge dynamics on the
hidden D7 branes in this case.

A more technical result concerns the actual calculability of soft
supersymmetry breaking terms. This has proved difficult in the
standard KKLT scenario due to the generic appearance of
tachyons. These are all absent in our models. There are several
scenarios for realistic soft breaking terms, depending on
the string scale and the location of the standard model fields on either
D3 or D7 branes. For D3 branes the bulk-induced soft terms are
hierarchically smaller than the gravitino mass. This realises several
proposals in the past (for  recent discussons of this class of models
see
\cite{kl,mazumdar}). For D7 branes the dominant
contribution comes from the flux induced soft terms which are all of
$\mc{O}(m_{3/2})$. The difference with previous results in the
no-scale approximation is that here the volume is fixed dynamically. 

It is interesting that intermediate string scales ($m_s\sim
10^{10-12}$ GeV ) are naturally preferred to obtain TeV scale soft terms and
stabilise the Higgs mass. The phenomenological virtues of these
scales were discussed in \cite{benakli, hepph9810535}. Nevertheless a
GUT scale scenario may give rise to the right scale of soft breaking
terms by including, for example, warping effects. Furthermore, even
the TeV scenario may be realised if the standard model is inside an
anti-brane or feels supersymmetry breaking directly. Here, as 
pointed out in section 4, the problem with the small mass of the
volume modulus would have to be addressed.

This opens new avenues for
phenomenological studies of realistic string models.
In particular, the bottom-up approach to string model building
proposed in \cite{hepth0005067} fits well with our scenario. The recent
models constructed in warped Calabi-Yau compactifications naturally
realise the Standard Model in our scenario.
It would be
interesting to explicitly  combine our results with realistic models such as
in {\cite{hepth0005067,hepth0312051,hepth0503079, hepth0408059, hepth0409132}}
and to explore further the
phenomenology of the different scenarios defined by the value of
the string scale. Furthermore, the structure of soft breaking terms we
have found has to  be complemented, in the large string scale scenarios,
with a low-energy analysis using the renormalisation group equations
to extrapolate the low-energy implications of our soft terms.
A recent analysis in this direction, using only the flux-induced
supersymmetry breaking, can be found in \cite{abi}.

There may be interesting cosmological implications.
In all cases there exist extremely flat directions, such as the one
corresponding to the volume axion. This is also protected by a
Peccei-Quinn symmetry making it stable under quantum corrections.
Ironically, in most cases this field is too light to be a candidate
for a quintessence field. We may ask the question backwards and look for
the value of the string scale such that this field has a mass of order
$10^{-33}$ eV. While there is some dependence on the associated gauge
group rank, this is typically close to the Planck scale ($m_s \sim 10^{17-18}
$ GeV.).

A further cosmological application could be to use the volume
modulus as an inflaton. Note that the potential is rather flat in
the large volume limit. In particular, in the regime where our
minimum coexists with the KKLT minimum, a saddle point develops
separating both minima, which may eventually give rise to
inflation. In particular, recent models of inflation in the KKLT
scenario either coming from
brane/antibrane systems \cite{kklmmt, bmnqrz, dt, kallosh1, tye, bcsq,
trivedi, hepth0501184} or the racetrack inflation \cite{racetrackinflation, rosssarkar}, could be
reconsidered in view of our results.

We must also address the overshooting problem
\cite{hepth9212049} as well as the cosmological moduli problem
\cite{hepph9308292, hepph9308325, ross}.
Note that in the intermediate scale scenario,
the volume modulus $T_5$ is on the border of being ruled out
by this problem.

There are also other open questions. The lack of control of the
value of the D7 moduli is probably the less understood part of our
construction. This is in general not a serious drawback. We know that
these moduli have compact support, and therefore must have a
minimum.
Recently it has been found using $F$-theory
techniques that D7 moduli tend not to be stabilised at the
orientifold limit \cite{hepth0501139}. It would then be interesting to
work in the F-theory setting with full control over the D7 moduli.

A further study of loop corrections, both from string theory and
field theory is desirable to pin down the magnitude of any effects
not included here. As described in section 3, these are principally
loop effects in the open string sector.
D3 brane loop corrections to the K\"ahler
potential are currently being computed in \cite{bhk}, and these may compete with
the $\alpha'^3$ corrections in the very large volume limit (see also
\cite{hepth9906039}). In the
regime of small coupling and large volume our solution will still
be dominant but in other regimes loop corrections may be important.
This may still be desirable depending on the sign of the
correction. If positive, it would provide a much better lifting
mechanism than any considered so far. If negative, it may
drive the potential to vanish from below at infinity with
$\alpha'$ corrections dominating at smaller (but still large)
volumes, still implying the existence of a large-volume minimum.
In any case we expect that our main results will be mostly
maintained but it would certainly be very interesting to have control
of all loop corrections to the K\"ahler potential.

Finally, loop corrections from the effective four-dimensional theory
may modify the values of the masses we have found. This depends
on the magnitude of the couplings. A preliminary analysis shows that
these corrections are suppressed by the volume and therefore do not tend
to destabilise the masses found above.\footnote{We thank
  C.P. Burgess for this estimate.} Finally, it would be desirable
to have a better control of the effects of warping. 
Although significant warping is not necessary to obtain a hierarchy
in the above models, 
it is very interesting that flux compactifications can realise both
the Randall-Sundrum scenario \cite{rs} with exponentially large geometric
hierarchies in the complex structure sector of the theory, whereas the
K\"ahler sector realises the large extra dimensions
scenario.
Exploring both these effects in
explicit string theoretical scenarios will certainly provide very
interesting phenomenology.

\section*{Acknowledgments}

We are grateful to B. Allanach, V. Balasubramanian, D. Baumann, P.
Berglund, R. Blumenhagen, C.P. Burgess, D. Cremades, G. Curio, B. de Carlos, F. Denef, B. Florea,
A. Font, 
M. B. Green, T. House, S. Kachru, 
L. McAllister, E. Palti,
A. Sinha, S. Stieberger and A. Uranga for interesting discussions. We particularly thank P.
C\'amara and L.E. Ib\'a\~nez for early collaboration on  soft breaking
terms. FQ thanks J.J. Blanco-Pillado for a conversation in 2003  
regarding the relative importance of perturbative and non-perturbative
effects in supergravity Lagrangians. 
We would also like to thank the anonymous referee for a careful and thorough
reading of the manuscript.
JC is grateful to EPSRC for a research studentship. FQ is
partially funded by PPARC and a Royal Society Wolfson award. KS would
like to thank Trinity College, Cambridge for financial support.

\begin{appendix}

\section{Dimensional Reduction}

The bosonic type IIB supergravity action in string frame is \cite{Polchinski}
\be
S_{IIB} = \frac{1}{(2 \pi)^7 \alpha'^4} \int d^{10}x \sqrt{-g}
\left\{ e^{-2 \phi}[\mc{R} + 4(\nabla \phi)^2] - \frac{F_1^2}{2} -
  \frac{1}{2 \cdot 3!} G_3 \cdot \bar{G}_3 - \frac{\tilde{F}_5^2}{4
    \cdot 5!} \right\}.
\ee
It is convenient to work in Einstein frame. We redefine
\be
\label{Einsteinframe}
g_{MN} = e^{(\phi - \phi_0)/2} \tilde{g}_{MN},
\ee
where $\phi_0$ =  $\langle \phi \rangle$. The factor of $e^{\frac{\phi_0}{2}}$ ensures that
$g$ and $\tilde{g}$ are identical in the physical vacuum.
The action is then
\be
\label{dilatonomitted}
\frac{2 \pi e^{-2 \phi_0}}{l_s^8} \int d^{10} x \sqrt{-\tilde{g}}
\left\{ \tilde{\mc{R}} - \frac{\partial_M S \partial^M \bar{S}}{2
  (\textrm{Re } S)^2}
- \frac{e^{\phi_0} G_3 \cdot \bar{G}_3}{12 \, \textrm{Re } S} - 
\frac{e^{2 \phi_0} \tilde{F}_5^2}{4 \cdot 5!} \right\},
\ee
where $l_s = 2 \pi \sqrt{\alpha'}$ and $S = e^{-\phi} + i C_0$.
We neglect warping effects; as discussed in section \ref{WarpingSection}, these are
subleading at large volume.
In the orientifold limit and in the absence of warping, $\tilde{F}_5 =
0$ and $\partial_M S = 0$.
The dimensional reduction of (\ref{dilatonomitted}) then gives
\be
S = \frac{2 \pi}{g_s^2 l_s^8} \left( \int d^{4} x \sqrt{-\tilde{g_4}} \tilde{\mc{R}}_4 \mc{V}
-  \overbrace{\int d^4 x \sqrt{-\tilde{g_4}} \left( \int d^6 x \sqrt{\tilde{g}_6}
\frac{e^{\phi_0} G_3 \cdot \bar{G}_3}{12 \, \textrm{Re } S}
\right)}^{V_{flux}} \right),
\ee
where $g_s = e^{\phi_0}$ and $\mc{V} = \int d^6 x \sqrt{\tilde{g}_6}$.
In 4d Einstein frame, the Einstein-Hilbert action is
\be
S_{EH} = \frac{1}{16 \pi G} \int d^4 x \sqrt{-g_E} \mc{R}_E \equiv
\frac{M_P^2}{2} \int d^4 x \sqrt{-g_E} \mc{R}_E.
\ee
$\tilde{g_4}$ and $g_E$ are related by $\tilde{g_4} = g_E \frac{{\cal V}^0_s}{\mc{V}_s}$,
where $\mc{V} \equiv \mc{V}_s l_s^6$ and ${\cal V}^0_s = \langle \mc{V}_s \rangle$.
This gives
\be
M_P^2 = \frac{4 \pi {\cal V}_s^0}{g_s^2 l_s^2} \quad \textrm{ and } \quad
m_s = \frac{g_s}{\sqrt{4 \pi {\cal V}_s^0}} M_P.
\ee

The K\"ahler potential is determined to be
\be
\label{FullKahlerPotential}
\frac{\mc{K}}{M_P^2} = - 2 \ln(\mc{V}_s) - 
\ln( S + \bar{S}) - \ln \left( -i \int \Omega \wedge \bar{\Omega} \right).
\ee
The superpotential can be found from $V_{flux}$.
If
\be
W = \frac{1}{l_s^2} \int G_3 \wedge \Omega,
\ee
then
\be
V_{flux} = \frac{4 \pi (\mc{V}_s^0)^2}{g_s l_s^4} \int d^4x 
\sqrt{-g_E} e^{\mc{K}} \left[ G^{a \bar{b}} D_a W D_{\bar{b}} W
- 3 W \bar{W} \right],
\ee
with $a, b$ running over all moduli.
Thus if
\be
\hat{W} = \frac{g_s^{\frac{3}{2}}M_P^3}{\sqrt{4 \pi} l_s^2} \int G_3 \wedge \Omega,
\ee
the scalar potential takes the standard $\mc{N} = 1$ form
\be
\label{scalarpotential}
V = \int d^4 x \sqrt{-g_E} e^{\mc{K}/M_P^2} \left[ G^{a \bar{b}}
D_a \hat{W} D_{\bar{b}} \bar{\hat{W}} - \frac{3}{M_P^2} \hat{W} \bar{\hat{W}}  \right].
\ee

The K\"ahler potential will receive perturbative corrections and the superpotential
non-perturbative corrections. The $\alpha'$ corrections to the K\"ahler potential
modify the 4-dimensional kinetic terms and arise from the
higher-derivative terms in the ten-dimensional IIB action.
The dimensional reduction of these gives the perturbative correction to the K\"ahler
potential. We then obtain
\be
\frac{\mc{K}}{M_P^2} = - 2 \ln(\mc{V}_s + \frac{\xi g_s^{3/2}}{2 e^{3 \phi/2}}) - \ln( S + \bar{S}) -
\ln \left( -i \int \Omega \wedge \bar{\Omega} \right).
\ee
where $\xi = - \frac{ \chi(M) \zeta(3)}{2 (2 \pi)^3}$.
The superpotential can receive non-perturbative corrections causing it to depend on
the K\"ahler moduli. These can arise from D3-brane instantons or gaugino condensation.
The generic form of the superpotential is then
\be
\hat{W} = \frac{g_s^{3/2} M_P^3}{\sqrt{4 \pi}} \left( W_0 + \sum
A_i e^{-a_i T_i} \right).
\ee

\section{Canonical Normalisation}

We here discuss the canonical normalisation of the K\"ahler moduli for
the 2-modulus example.
The K\"ahler potential is given by
\be
\label{KahlerPotential}
\mc{K} = \mc{K}_{cs} - 2 \ln \left(\left(T_5 + \bar{T}_5)\right)^{\frac{3}{2}} -
\left(T_4 + \bar{T}_4\right)^{\frac{3}{2}}\right) + \textrm{ constant }.
\ee
Here $iT_4 = b_4 + i\tau_4$ and $iT_5 = b_5 + i \tau_5$.
All terms depending on dilaton and complex structure moduli
have been absorbed into $\mc{K}_{cs}$. We
also recall that as $\tau_5 \gg 1$,
$\frac{1}{\tau_5}$ serves as a good expansion parameter.

It may be verified that
\bea
\partial_{T_5} \partial_{\bar{T}_5} K & = & \frac{3 \tau_5^{-\half}}{4(\tau_5^{\frac{3}{2}}
- \tau_4^{\frac{3}{2}})} + \frac{9 \tau_4^{\frac{3}{2}} \tau_5^{-\half}}{8 (\tau_5^{\frac{3}{2}} - \tau_4^{\frac{3}{2}})^2},
\nonumber \\
\partial_{T_4} \partial_{\bar{T_5}} K = \partial_{T_5} \partial_{\bar{T_4}} K & = &
\frac{-9 \tau_4^{\frac{1}{2}} \tau_5^{\half}}{8 (\tau_5^{\frac{3}{2}} -  \tau_4^{\frac{3}{2}})},  \nonumber \\
\partial_{T_4} \partial_{\bar{T_4}} K & = & \frac{3 \tau_4^{-\half}}{4(\tau_5^{\frac{3}{2}}
- \tau_4^{\frac{3}{2}})} + \frac{9 \tau_4}{8 (\tau_5^{\frac{3}{2}} - \tau_4^{\frac{3}{2}})^2}.
\nonumber
\eea
These results are summarised by
\be
G = \left( \begin{array}{cc} G_{4\bar{4}} & G_{4\bar{5}} \\
G_{5\bar{4}} & G_{5\bar{5}} \end{array} \right) =
\left( \begin{array}{cc} \frac{3 \tau_4^{-\half}}{8\tau_5^{\frac{3}{2}}} + \mc{O}(\frac{1}{\tau_5^3}) &
\frac{-9 \tau_4^\half}{8 \tau_5^{\frac{5}{2}}} + \mc{O}(\frac{1}{\tau_5^4}) \\
\frac{-9 \tau_4^\half}{8 \tau_5^{\frac{5}{2}}} + \mc{O}(\frac{1}{\tau_5^4}) &
\frac{3}{4\tau_5^{2}} + \mc{O}(\frac{1}{\tau_5^{\frac{7}{2}}}) \end{array} \right).
\ee
We denote the values of the fields $(b_4, b_5,
\tau_4, \tau_5)$ at the minimum by $(b_4^0, b_5^0, \tau_4^0,
\tau_5^0)$ and now define
$$
\tau_5^{'} = \sqrt{\frac{3}{2}} \frac{\tau_5}{\tau_5^0}, \quad \tau_4^{'} =
\sqrt{\frac{3}{4}} \frac{\tau_4}{(\tau_5^0)^{\frac{3}{4}}(\tau_4^0)^{\frac{1}{4}}},
\quad b_5^{'} = \sqrt{\frac{3}{2}} \frac{b_5}{\tau_5^0}, \quad b_4^{'} =
\sqrt{\frac{3}{4}} \frac{b_4}{(\tau_5^0)^{\frac{3}{4}}(\tau_4^0)^{\frac{1}{4}}},
$$
$$
\textrm{ and } \quad \tau_4^{'0} =
\sqrt{\frac{3}{4}} \frac{\tau_4^0}{(\tau_5^0)^{\frac{3}{4}}(\tau_4^0)^{\frac{1}{4}}}, \quad
  \tau_5^{'0} = \sqrt{\frac{3}{2}}.
$$
Then after some manipulation, we obtain
\bea
\lefteqn{\sum_{i,j} G_{i \bar{j}} \partial_\mu T^i \partial^\mu
  \bar{T}^j = } \nonumber \\
& & \half \Bigg[ \left( \frac{\tau_5^{'0}}{\tau_5^{'}} \right)^{\frac{3}{2}} \left( \frac{\tau_4^{'0}}{\tau_4^{'}}
\right)^{\frac{1}{2}} \partial_\mu b_4^{'} \partial^\mu b_4^{'} + \left(
\frac{\tau_5^{'0}}{\tau_5^{'}} \right)^2 \partial_\mu b_5^{'}
\partial^\mu b_5^{'} - 4\sqrt{3} \left( \frac{\tau_4^{'}}{\tau_4^{'0}}\right)^\half
\left(\frac{\tau_5^{'0}}{\tau_5^{'}}\right)^{\frac{5}{2}} \tau_4^{'0}
\partial_\mu b_4^{'} \partial^\mu b_5^{'}  \nonumber \\
& & + \left( \frac{\tau_5^{'0}}{\tau_5^{'}} \right)^{\frac{3}{2}} \left( \frac{\tau_4^{'0}}{\tau_4^{'}} \right)^{\half}
\partial_\mu \tau_4^{'} \partial^\mu \tau_4^{'} + \left(
\frac{\tau_5^{'0}}{\tau_5^{'}} \right)^2 \partial_\mu \tau_5^{'}
\partial^\mu \tau_5^{'} -4\sqrt{3} \left( \frac{\tau_4^{'}}{\tau_4^{'0}}\right)^\half
\left(\frac{\tau_5^{'0}}{\tau_5^{'}}\right)^{\frac{5}{2}} \tau_4^{'0}
\partial_\tau \tau_4^{'} \partial^\tau \tau_5^{'} \Bigg] \nonumber
\eea
The moduli are now canonically normalised except for the crossterm,
which is suppressed by $(\tau_5^0)^{\frac{3}{4}}$ and is thus
very small. All such crossterms could be eliminated by field
redefinitions order by order in $\frac{1}{\tau_5}$; however,
negelecting the subleading corrections it is sufficient to use
$\tau_5^{'}$ and $\tau_4^{'}$ as canonically normalised fields.

\section{Soft Terms}

Let us introduce the notation ${\cal{V}}'=(T_5+\bar{T}_5)^{3/2}-(T_4+\bar{T}_4)^{3/2}$
(which differs by a factor of $36$ from the Calabi-Yau volume $\cal{V}$).
We write
\begin{eqnarray}
\hat{\K}_{T_5} = -3{(T_5+\bar{T}_5)^{1/2}\over{{\cal{V}}'+{\xi'/2}}},\nonumber\\
\hat{\K}_{T_4} = 3{(T_4+\bar{T}_4)^{1/2}\over{{\cal{V}}'+{\xi'/2}}}.
\end{eqnarray}
We denote $(T_4+\bar{T}_4)^{1/2} = X$ \footnote{We do not simply use
  $\tau_4$ as this would not be valid when $T_4$ includes D3-brane
  moduli.} and  $(T_5+\bar{T}_5)^{1/2}=Y$.
Then ${\cal{V}}' = Y^3-X^3$ and we can calculate the metric $\K_{T_i \bar{T}_j}$:
\begin{equation}
\K_{T_i \bar{T}_j}=\left(
\begin{array}{cc}
{3\over{2Xx}}+{9X^2\over{2x^2}} &  -{{9XY}\over{2x^2}}\\
-{{9XY}\over{2x^2}} & {-3\over{2Yx}}+ {{9Y^2}\over{2x^2}}
\end{array}
\right),
\end{equation}
and the inverse metric $\K^{\bar{T}_i T_j}$
\begin{equation}
\left(
\begin{array}{cc}
-2X ({\cal{V}}'+{\xi/2})(2Y^3+X^3 -{\xi'/2})\over{3(\xi'/2-2{\cal{V}}')}
& -2X^2 Y^2{({\cal{V}}'+{\xi'/2})\over{({\xi'/2}-2{\cal{V}}')}}\\
-2X^2 Y^2{({\cal{V}}'+{\xi'/2})\over{({\xi'/2}-2{\cal{V}}')}}
& -2Y ({\cal{V}}'+{\xi'/2})(2X^3+Y^3 +{\xi'/2})\over{3(\xi'/2-2{\cal{V}}')}
\end{array}
\right).
\end{equation}
In these expressions $x = Y^3 - X^3+\xi'/2 = {\cal{V}}'+\xi'/2.$ We
can now calculate the F-terms as given by formula (\ref{fterms}).
We assume that $\partial_5 W\ll \partial_4 W$ (where $\partial_i\equiv
\partial_{T_i}$) and only include the
nonperturbative contribution corresponding to $T_4.$ The result is
\begin{eqnarray}
F^4 &=& e^{\hat{\K}/2}{{2{\cal{V}}'+\xi'}\over{2{\cal{V}}'-{\xi'/2}}} \left(
-X^2 W + {X\over3} (2Y^3+X^3 - {\xi\over2})(\partial_4 W)\right)\nonumber\\
F^5 &=& e^{\hat{\K}/2}{{2{\cal{V}}'+\xi}\over{2{\cal{V}}'-{\xi'/2}}}
\left( -Y^2 W + X^2 Y^2 (\partial_4 W))
\right)
\end{eqnarray}
After a redefinition of $c$ and $d$, the prefactor of the kinetic term for
the brane modulus $\phi$, $\tilde{\K}_i$,can be rewritten as
\begin{equation}
\tilde{\K}_i = {{cX+dY}\over{{\cal{V}}'+{\xi'/2}}}.
\end{equation}
To calculate scalar masses we need $\partial_m \partial_{\bar{n}}
\log{\tilde{\K}_i} = \partial_m \partial_{\bar{n}}(\log (cX+dY)
- \log({\cal{V}}'+{\xi'\over2})).$ As $\hat{\K} =
-2\log({\cal{V}}'+\xi'/2)+{\rm const.}$, 
$ - \partial_m \partial_{\bar{n}} \log ({\cal{V}}'+{\xi'\over2})
 = (1/2) \hat{\K}_{m\bar{n}}$ and so
\begin{equation}
\partial_m \partial_{\bar{n}}
\log{\tilde{\K}_i} = \partial_m \partial_{\bar{n}}(\log (cX+dY))
+ {1\over2}\hat{\K}_{m\bar{n}}.
\end{equation}
The necessary derivatives of $\log (cX+dY)$ can be calculated using
$\partial_4 X = 1/(2X), \partial_5 Y = 1/(2Y)$, and likewise
with respect to $\bar{4}$ and $\bar{5}.$ The results are
\begin{eqnarray}
\label{logcu}
\partial_4 \partial_{\bar{4}}
\log (cX+dY) &=& -{c (2cX+dY)\over{4X^3(cX+dY)^2}},\nonumber\\
\partial_4 \partial_{\bar{5}} \log (cX+dY) &=& -{cd\over{4XY(cX+dY)^2}},
\nonumber\\
\partial_5 \partial_{\bar{5}}
\log (cX+dY) &=& -{d(2dY+cX)\over{4Y^3 (cX+dY)^2}}.
\end{eqnarray}

We now wish to find the soft masses, expressing the result as a small
deviation from the no-scale result, which is zero. To do this we 
make an expansion assuming that $(\partial_4 W)$ is small (we know
it is  $\mc{O}(1/{\cal{V}})$ at the minimum of the scalar potential) and
$\xi/{\cal{V}}$ is small. For example, we can note that $F^m \bar{F}^{\bar{n}}$ always
has a prefactor of $(2{\cal{V}}'+\xi')^2/{(2{\cal{V}}'-{\xi'/2})^2}$
which may be expanded as
\begin{equation}
1+{{3\xi'}\over{2\cal{V}}'} + {\cal{O}}\left( 1\over{\cal{V}}^2 \right).
\end{equation}
From the above analysis we expect that
$F^m \bar{F}^{\bar{n}} \partial_m \partial_{\bar{n}}
\log (cX+dY)$ can be written as $e^{\hat{\K}} ((-1/2)|W|^2 + {\cal{O}}
(1/{\cal{V}}^\alpha))$ for some $\alpha>0.$
Indeed, an explicit computation gives
\begin{eqnarray}
\lefteqn{F^m \bar{F}^{\bar{n}} \partial_m \partial_{\bar{n}} \log (cX+dY) =
e^{\hat{\K}}
\left( 1+{{3\xi'}\over{2\cal{V}}'}\right) \ti} \nonumber\\
& &\left(
-{1\over2}|W|^2 + {1\over{3(cX+dY)}} \left( -{{\xi c}\over2}
+ 3d X^2 Y + c X^3 + 2 c Y^3\right)W (\partial_4 W) \right). \nonumber
\end{eqnarray}
Similarly we have
\begin{equation}
F^m \bar{F}^{\bar{n}} \hat{\K}_{m\bar{n}} =
e^{\hat{\K}} {{3{\cal{V}}'(2{\cal{V}}'-{\xi'\over2})\over{2({\cal{V}}'+{\xi'\over2})^2}}}
\left( |W|^2 - {W(\partial_4 W)\over{3\cal{V}}'} \left(
2X^2 {\xi'\over2} + 2X^2 {\cal{V}}') \right)\right).
\end{equation}
Expanding ${\cal{V}}'(2{\cal{V}}' - {\xi'/2})/(2({\cal{V}}'+{\xi'/2})^2)$
as $1-3\xi'/(4{\cal{V}}')$ we get
\begin{eqnarray}
F^m \bar{F}^{\bar{n}} \partial_m \partial_{\bar{n}} \log {\tilde{\K}_i} &=&
e^{\hat{\K}} \left( 1+{{3\xi'}\over{2\cal{V}}'} \right) \Bigl[
-|W|^2  \nonumber\\
& &
+{1\over{3(cX+dY)}} \left( -{{\xi' c}\over2}+3d X^2 Y + cX^3 + 2c Y^3
\right) W(\partial_4 W) \nonumber\\
& & -W(\partial_4 W) X^2 (1+ {\xi'\over{2\cal{V}}'})\Bigr]
\end{eqnarray}
Further simplifying and neglecting ${\cal{O}} (1/{\cal{V}}^2)$ terms
we obtain
\begin{eqnarray}
\lefteqn{e^{\hat{\K}} ( -|W|^2 + {{3\xi'}\over{2\cal{V}}'}|W|^2} \nonumber\\
& &+ {W\over{3(cX+dY)}} \left(
{{-\xi' c}\over 2} + 3d X^2 Y + cX^3 + 2cY^3 \right) (\partial_4 W)
- W(\partial_4 W) X^2). \nonumber
\end{eqnarray}
The $-e^{\hat{\K}} |W|^2$ cancels with $m_{3/2}^2$ so the mass squared of the brane
modulus can be seen to have the form
\begin{equation}
\label{smallmass}
{{3\xi'}\over{2\cal{V}}'} + f(X,Y) (\partial_4 W),
\end{equation}
which is what we expect naively --- before the no-scale structure is
broken, the mass is simply zero, and there are two sources of no-scale
breaking: $\alpha'$ corrections corresponding to the first term in
(\ref{smallmass}) and nonperturbative corrections corresponding to
the second term. To estimate the size of the soft terms, note that
all the constants involved in (\ref{smallmass}) are expected to be
${\cal{O}} (1)$ (including $c, d, \xi'$) and so the first term
is ${\cal{O}}(1/\cal{V}).$ As the volume is
${\cal{V}}\sim Y^3$, we can estimate
\begin{equation}
f(u,v)\sim {Y^3\over Y} = Y^2 \sim {\cal{V}}^{2/3}.
\end{equation}
We also note that, at the minimum, $\partial_4 W \sim \frac{W_0}{\mc{V}}$.
Then the volume scaling of the second term is ${\cal{V}}^{2/3}/{\cal{V}}
={\cal{V}}^{-1/3}$, and
as $e^{\hat{\K}} \sim 1/{{\cal{V}}^2}$ the moduli masses squared
scale as ${\cal{O}} (1/{\cal{V}}^{7/3}).$
Finally, by considering the form of the superpotential
(\ref{wnodilaton}), we see that $W (\partial_4 W) \sim W_0^2 g_s^4$.
Putting the factors together, we get
\be
m_i^2 = \mc{O}(1) \frac{g_s^4 W_0^2}{4 \pi (\mc{V}_s^0)^{7/3}} M_P^2.
\ee

\end{appendix}

\end{document}